\documentclass[a4paper]{aa} 
\usepackage{graphicx} 
\usepackage{times}
\usepackage{natbib} 
\bibpunct{(}{)}{;}{a}{}{,} % to follow the A&A style 

\newcommand{\msun}{$\mathrm{{M}_\odot}$}

\newcommand{\mjup}{$\mathrm{{M}_J}$}
\newcommand{\mjups}{$\mathrm{{M}_J}$ }

\newcommand{\mearth}{$\mathrm{{M}_\oplus}$}
\newcommand{\mearths}{$\mathrm{{M}_\oplus}$ }

\authorrunning{Paardekooper \& Mellema}

\newcommand{\eqref}[1]{(\ref{#1})}

\begin{document}
  \title{Growing and moving low-mass planets in non-isothermal disks} 
  \author{Sijme-Jan Paardekooper \and Garrelt Mellema} 
  \author{Sijme-Jan Paardekooper \inst{1,2} \and Garrelt Mellema \inst{3,2}} 
  \offprints{S. J. Paardekooper\\\email{S.Paardekooper@damtp.cam.ac.uk}} 
  \institute{Department of Applied Mathematics and Theoretical Physics, University of Cambridge, Wilberforce Road, Cambridge CB3 0WA, UK\\
             \email{S.Paardekooper@damtp.cam.ac.uk} \and
             Leiden Observatory, Leiden University, Postbus 9513, NL-2300 RA Leiden, 
             The Netherlands \and           
             Stockholm Observatory, AlbaNova University Center,
	   Stockholm University, SE-106 91 Stockholm, Sweden \\
	   \email{garrelt@astro.su.se}}

  \date{Draft Version \today} % (Received ; Accepted)}
  
  \abstract{}{We study the interaction of a low-mass planet with a 
    protoplanetary disk with a realistic treatment of the energy balance by 
    doing radiation-hydrodynamical simulations. We look at accretion and
    migration rates and compare them to isothermal studies.}
    {We used a three-dimensional version of the hydrodynamical method RODEO,
    together with radiative transport in the flux-limited diffusion approach.}
    {The accretion rate, as well as the torque on the planet, depend critically on the ability of the
    disk to cool efficiently. For densities appropriate to 5 AU in the
    solar nebula, the accretion rate drops by more than an order of magnitude compared to isothermal models, while at the same time the torque on the planet is \emph{positive}, indicating outward migration.
    It is necessary to lower the density by a factor of 2 to recover
    inward migration and more than 2 orders of magnitude to recover the usual 
    Type I migration. The torque appears to be proportional to the radial entropy gradient in the unperturbed disk. These findings are critical for the survival of protoplanets, and they should ultimately find their way into population synthesis models. }{} 
    
  
  \maketitle
  
\section{Introduction}
Planet formation takes place in protoplanetary disks that are commonly found
around young stars \citep{1996Natur.383..139B}. These disks play a critical 
role in structuring the forming planetary system in terms of masses and orbits,
and planet-disk interaction is therefore a very important process to study
in order to understand planet formation. 

Planet-disk interaction affects a forming planet in three ways. 
First of all, the mass that the disk is able to supply to the young planet
is limited, which leads to a maximum gas mass that a planet can achieve
\citep{1999ApJ...526.1001L,1999MNRAS.303..696K,2003ApJ...586..540D,
2003MNRAS.341..213B}. This limiting mass is of the order of a few Jupiter 
masses (\mjup). But also for planets of lower mass the disk regulates the mass 
accretion, which becomes especially apparent in three-dimensional numerical 
simulations \citep{2003ApJ...586..540D}. 

Second, the disk is able to change the orbital radius of the planet through 
tidal interaction \citep{1980ApJ...241..425G}. Depending on the mass of the
planet, two modes of migration can be distinguished 
\citep{1997Icar..126..261W}: type I migration for low-mass planets that 
generate a linear disk response, and type II for high-mass, gap-opening 
planets. Recently a third type of migration was put forward, in which strong
corotational torques force a very fast mode of migration 
\citep{2003ApJ...588..494M,2004ASPC..324...39A}. This Type III migration
is very sensitive to local density gradients, in terms of migration speed
and even migration direction, and therefore the structure of the gas disk 
determines the outcome of the migration process. This is unlike Type I and 
Type II migration, for which all planets move inward and only the time scale
varies with disk properties.

Finally, the eccentricity of the planet's orbit may be affected by the disk. 
Usually, eccentricity is damped \citep{1993ApJ...419..166A}, but high-mass
planets ($M_\mathrm{p} > 5 $ \mjup) that open deep gaps may experience 
eccentricity growth \citep{2001A&A...366..263P,2006A&A...447..369K}. However, 
due to possible saturation of corotation resonances planets of $M_\mathrm{p}
\approx$ \mjups may pick up significant eccentricity during the gap formation 
process \citep{2004ApJ...606L..77S}, although this requires a slightly 
eccentric orbit to start with \citep{2003ApJ...585.1024G}.

Observations of the masses and orbits of extrasolar planets provide insight
in the result of planet-disk interaction. One of the most striking results
that were obtained is the discovery of a new class of planets, the so-called
Hot Jupiters. They are gas giant planets orbiting very close to their central
star. It is unlikely that they have formed in situ \citep{1999ApJ...521..823P}
and inward migration through planet-disk tidal interaction has been put forward
as the explanation for their existence. 

The main problem for this scenario is that the time scales for migration as
obtained from analytical \citep{2002ApJ...565.1257T} and numerical 
\citep{2003MNRAS.341..213B,2003ApJ...586..540D} arguments are
usually smaller than or comparable to the total lifetime of the disk. This
means that theory essentially predicts that all planets should fall onto the
central star, which poses a problem for planet formation theory. If it is
already hard to form a single giant planet within the lifetime of the disk
\citep{1996Icar..124...62P}, how are we going to form a huge amount of them,
of which only a tiny fraction survives? Clearly a stopping mechanism is 
required that prevents planets from migrating all the way towards the star.

One possibility is that the planet encounters an inner hole in the disk,
possibly created by the magnetic field of the central star. When there is no 
disk around to interact with, the planet will stop migrating. Another
possibility is that the planet undergoes Type III outward migration at 
some point \citep{2003ApJ...588..494M}. Finally, direct interaction of the
disk with a magnetic field may stop the planet, either by a toroidal
magnetic field \citep{2003MNRAS.341.1157T} or, for a low-mass planet, its migration may be slowed down through magnetic turbulence \citep{2004MNRAS.350..849N}.

However, most numerical hydrodynamical work on planet-disk interaction lacks
proper treatment of the energy balance, with only a few notable exceptions
\citep{2003ApJ...599..548D,2006A&A...445..747K,2003MNRAS.346..915M}. The 
usually adopted locally isothermal equation of state assumes that all excess 
energy generated by compression, viscous dissipation or shocks can be radiated 
away efficiently, thereby keeping the temperature profile fixed. However, the 
ability of the disk to cool is strongly linked to the opacity, and therefore 
to the density.

The effects of temperature structure on planet migration in realistic disks
were studied by \cite{2003ApJ...593.1116J,2004ApJ...608..497J,
2005ApJ...619.1123J} who used a detailed disk model including radiative 
transfer to calculate the torque on the planet in an analytical way. They
found that temperature perturbations resulting from shadowing and illumination
at the surface of the disk can decrease the migration rate by a factor of
2. \cite{2004ApJ...606..520M} also find reduced migration rates in realistic
T Tauri $\alpha$-disks, and they point out the importance of sudden changes
in the opacity, for example near the snow line.

In this study, we aim at relaxing the isothermal assumption in hydrodynamical
simulations of planet-disk interaction. The first results on planet
migration were presented in \cite{radlett}, where it was shown that
low-mass planets may migrate \emph{outward} in disks with realistic
temperature structures. In this paper, we continue our study of
migration of low-mass objects, with more eye for detail than in
\cite{radlett}, and include results on accretion. 

In contrast with 
\cite{2006A&A...445..747K}, we focus on low-mass planets that do not open
a gap in the disk. For these deeply embedded planets temperature effects
are the most important. However, due to their low mass their gravitational
sphere of influence is small, and one needs very high resolution locally 
to resolve the atmosphere of the planet. Furthermore, because radiation is 
the main cooling agent, we need to perform radiation-hydrodynamical 
simulations. This is a major leap from isothermal simulations, because 
we are not only adding an equation for the gas energy but also the evolution
of the radiation field needs to be followed. 
 
The plan of this paper is as follows. We briefly review the relevant cooling
time scale in Sect. \ref{3DsecCool}, effects of radiation in
Sect. \ref{3DsecRad} and the condition for convection in
Sect. \ref{3DsecConv}. In Sect. \ref{3DsecModel} 
we present the adopted disk model, and in Sect. \ref{3DsecNum} we discuss the 
numerical method. In Sect. \ref{3DsecRes} we present the results, from simple 
isothermal models to the full radiation-hydrodynamical models. We give a 
discussion on the results in Sect. \ref{3DsecDisc}, and we conclude in Sect. 
\ref{3DsecCon}.
 
\section{Cooling properties}
\label{3DsecCool}
We take the analytical opacity data from \cite{1994ApJ...427..987B}, which
state that for low temperatures (beyond the snow line) the Rosseland mean 
opacity in $\mathrm{cm^{-1}}$ is:
\begin{equation}
\label{3DeqOpacLow}
\kappa=\kappa_0 \rho T^2,
\end{equation}
where $\rho$ is the density in $\mathrm{g~cm^{-3}}$, $T$ denotes the 
temperature (K) and $\kappa_0=2.0~10^{-4}$ is a constant. We discuss the 
cooling properties of the disk as a function of opacity, starting with the
optically thin regime.

The time scale for the coupling between gas and radiation is given by
\begin{equation}
t_\mathrm{coupling}=\frac{1}{\kappa c},
\end{equation}
where $c$ is the speed of light in vacuum. When $t_\mathrm{coupling}$ is
larger than the dynamical time scale $t_\mathrm{dyn}$, which equals the
orbital time scale of the planet, the gas can not transfer its internal energy
to the radiation field and therefore the gas will not cool efficiently through
radiation. Plugging in the opacity of Eq. \eqref{3DeqOpacLow}, and assuming 
$T\approx 50$ K, which is appropriate for the location of Jupiter in the solar nebula, we find that the density near the planet should satisfy
\begin{equation}
\label{3DeqCoupling}
\rho > 10^{-18}~\mathrm{g~cm^{-3}},
\end{equation}
in order to cool through radiation. At the midplane of a protoplanetary disk
in the planet forming region this condition is always satisfied.

In general, when a region is optically thin and Eq. \eqref{3DeqCoupling} holds 
it can cool efficiently because all its thermally emitted photons can escape 
from this region. In a protoplanetary disk the scale length of interest is the 
disk thickness $H$. For planet-disk interaction specifically, most of the 
torque on the planet comes from material that is closer than $\sim 2 H$ from 
the planet \citep{2003MNRAS.341..213B}. The condition for the disk to be 
optically thin over one pressure scale height is given by:
\begin{equation}
\kappa H < 1.
\end{equation}
Putting in our opacity law Eq. \eqref{3DeqOpacLow} we can write a condition for
the density:
\begin{equation}
\label{3DeqCoolthin}
\rho < \frac{1}{\kappa_0~T^2~H}.
\end{equation}
At the location of Jupiter in a typical protoplanetary disk $T \approx 50$ K 
and $h\equiv H/r =0.05$, where $r$ is the orbital distance of Jupiter. Putting 
in these values, we find that for the disk to be optically thin over one 
pressure scale height $\rho < 5.1~10^{-13}~\mathrm{g~cm^{-3}}$, while the 
density at 5 AU in the minimum mass solar nebula (MMSN) is given by 
$\rho=10^{-11}~\mathrm{g~cm^{-3}}$. 

When a region is optically thick, it can still cool efficiently depending on 
the local conditions. Consider a sphere of radius $H$ around the position of 
the planet. The flux through this sphere in the optically thick case can be 
approximated by
\begin{equation}
F_\mathrm{R}=\frac{\sigma T^4}{\tau},
\end{equation}
where $\tau=\kappa H$ is the optical depth over the radius of the sphere and
$\sigma$ is the Stefan-Boltzmann constant. The internal energy density in the 
sphere is given by:
\begin{equation}
\epsilon=\frac{p}{\Gamma-1}=\frac{\rho \frac{{\mathrm R}}{\mu} T}{\Gamma-1}=
\frac{\tau \frac{{\mathrm R}}{\mu}}{\kappa_0 T H (\Gamma-1)},
\end{equation}
where $p$ denotes gas pressure, $\Gamma$ is the adiabatic exponent, R is the 
universal gas constant and $\mu$ is the mean molecular weight. The cooling 
time scale is given by the total internal energy content of the sphere 
divided by the total energy that flows through the surface of the sphere:
\begin{equation}
\label{3DeqCoolthick}
t_\mathrm{cool}=\frac{\epsilon~\frac{4}{3}\pi H^3}{F_\mathrm{R}~4\pi H^2}=
\frac{\epsilon H}{3F_\mathrm{R}}=\frac{\tau^2 \frac{{\mathrm R}}
{\mu}}{3(\Gamma-1)\kappa_0 \sigma T^5}.
\end{equation}
Putting in the density and temperature appropriate for the location of Jupiter
($\rho=10^{-11}~\mathrm{g~cm^{-3}}$, $T=50$ K and $\mu=2.4$) we find that 
$t_\mathrm{cool}\approx 10~t_\mathrm{dyn}$, where the dynamical time scale
$t_\mathrm{dyn}$ equals the orbital time scale of the planet. Cooling is 
only efficient when $t_\mathrm{cool}< t_\mathrm{dyn}$, and therefore we 
conclude that cooling is \emph{not} efficient for these parameters.

It is easy to show that the cooling time scale depends on the distance to
the central star through:
\begin{equation}
t_\mathrm{cool} \propto \frac{\rho^2(r)H^2(r)}{T(r)}.
\end{equation}
When we assume a constant aspect ratio $h$, which implies that the temperature
varies with radius as $r^{-1}$, and a power law for the density with index
$-3/2$ we see that the cooling time does not depend on radius. This means
that 
\begin{equation}
\label{3Deqtcool}
\frac{t_\mathrm{cool}}{t_\mathrm{dyn}}\approx 10 \left(\frac{r}{5~\mathrm{AU}}
\right)^{-\frac{3}{2}}.
\end{equation}
This means that at approximately 15 AU in the MMSN $t_\mathrm{cool} \approx
t_\mathrm{dyn}$. Inside this radius, cooling will not be efficient.

Both Eq. \eqref{3DeqCoolthin} and Eq. \eqref{3DeqCoolthick} have a strong 
temperature dependence, and therefore it is not possible to determine in 
advance if a disk with an embedded planet that changes the local temperature 
and density will be able to cool efficiently. However, this analysis shows that
for typical densities and temperatures including proper cooling mechanisms is
essential for planet-disk interaction.

Summarizing, for a fixed temperature of 50 K at $5$ AU, we can distinguish 
four cooling regimes:
\begin{itemize}
\item{For the lowest densities (see Eq. \eqref{3DeqCoupling}) the gas can not 
cool through radiation.}
\item{For somewhat higher densities, gas is able to cool through radiation but
the disk is still optically thin over a distance $H$.}
\item{Even higher densities make the disk optically thick over $H$, which means
that thermally emitted photons do not immediately leave the region that 
provides the torque on the planet.}
\item{For the highest densities the disk is optically thick and 
$t_\mathrm{cool}> t_\mathrm{dyn}$, which means that the disk can not cool 
efficiently.}
\end{itemize}
Note that only the second regime resembles the isothermal limit, while the 
first regime is not important for low-mass planets that remain deeply embedded.
In the third and the fourth regime, temperature effects may play a large role 
in determining the total torque onto the planet.

\section{Radiative effects}
\label{3DsecRad}
In general, radiation is not only a cooling agent but it may also dynamically
change the velocity structure around the planet. The effect of radiation on 
the gas velocity goes through radiation pressure, which, in the isotropic
case, is given by
\begin{equation}
P=\frac{1}{3} a T^4,
\end{equation}
where $a$ is the radiation constant. The gas pressure $p$ is again given by the
ideal gas law, and we can then write:
\begin{equation}
\frac{P}{p}=\frac{\frac{1}{3}a T^4}{\rho \frac{R}{\mu}T}=
\frac{\mu~a~T^3}{3\rho~R}.
\end{equation}
For the typical temperature at $5$ AU, $T=50$ K, we get:
\begin{equation}
\frac{P}{p}=\frac{1.0~\cdot~10^{-17}}{\rho}.
\end{equation}
For a typical midplane density of $\rho=10^{-11}~\mathrm{g~cm^{-3}}$ radiation
pressure is negligible compared to gas pressure.

This is an important result, because it shows that near a deeply embedded
planet the dynamical effects of radiation are not important. This means that 
the gas pressure and velocity structure will be almost the same as in the
locally isothermal case. This greatly simplifies the torque analysis: in 
regions where the temperature rises because of gas compression or the release 
of potential energy the density must be lower than in the isothermal case
to arrive at the same pressure. Therefore these regions will exert a lower
torque onto the planet. Any asymmetry in heating will therefore show up as
a torque asymmetry, which may alter the total torque balance on the planet
considerably.

\section{Convection}
\label{3DsecConv}
Heat transport in a planetary envelope can occur through radiation or through
convection. Convection sets in when the temperature gradient becomes too 
steep for radiation to transport energy towards the surface of the planet. 
As is well known from theory of stellar interiors, this happens when
\begin{equation}
\label{3DeqConv}
\frac{\partial T}{\partial{z}} < \left. \frac{\partial T}{\partial{z}}
\right|_\mathrm{ad},
\end{equation}
where $z$ is the direction towards the surface of the planet and the 
right hand side denotes the adiabatic temperature gradient, which can be
expressed as:
\begin{equation}
\label{3DeqDTad}
\left. \frac{\partial T}{\partial{z}}\right|_\mathrm{ad}=
-\frac{T}{H}~\frac{\Gamma}{\Gamma-1}.
\end{equation}
We can make a simple estimate of the importance of convective energy transport
by approximating the actual temperature gradient in the planetary envelope as:
\begin{equation}
\label{3DeqDT}
\frac{\partial T}{\partial{z}} \approx 
\frac{T_\mathrm{neb}-T_\mathrm{c}}{R_\mathrm{R}},
\end{equation}
where $T_\mathrm{neb}$ is the temperature of the surrounding nebula, 
$T_\mathrm{c}$ is the central temperature of the planet and $R_\mathrm{R}$ 
denotes the Roche lobe of the planet. The assumption is that the planet 
connects to the disk at a distance given by the Roche lobe, and that the disk
remains unperturbed from this distance. This is the usual assumption in
one-dimensional planet formation models \citep{1996Icar..124...62P}.

When we further assume that the low-mass planets do not change to local 
pressure scale height in the disk, we can plug Eq. \eqref{3DeqDTad} and Eq. 
\eqref{3DeqDT} into Eq. \eqref{3DeqConv} to obtain a maximum value for
the central temperature of the planet:
\begin{equation}
\label{3DeqTconv}
T_\mathrm{c} < \left(1+\frac{R_\mathrm{R}}{H}~\frac{\Gamma}{\Gamma-1}\right)
T_\mathrm{neb}.
\end{equation}
For a 5 \mearths planet embedded in a disk with aspect ratio $h=0.05$ and
adiabatic exponent $\Gamma=1.4$ we find that $T_\mathrm{c} < 2.21~
T_\mathrm{neb}$. 

As will become apparent in Sect. \ref{3DsecRes}, the direction of
migration critically depends on the radial entropy gradient in the
disk, or, in other words, whether Eq. \eqref{3DeqConv} is satisfied in
the radial direction. For a disk that follows a power law in
temperature and density with indices $\beta$ and $\alpha$,
respectively, the entropy also follows a power law with index
\begin{equation}
\label{eqent}
\mathcal{S}=\beta-(\Gamma-1)\alpha.
\end{equation}
Our nominal model, with $\beta=-1$ and $\alpha=-3/2$, has a negative
entropy gradient. Note that because of the rotation of the disk, this
does not lead to convection. Also, we did not find evidence for an
instability due to baroclinic effects \citep{2003ApJ...582..869K}.

\section{Model design}
\label{3DsecModel}
Our model consists of three objects: the central star, the planet and the disk.
Below, we give a short description of each of these components. Throughout
this paper, we will work in a spherical polar coordinate system 
$(r,\theta,\phi)$ that co-rotates with the planet.

\subsection{Central star}
The central star is the main source of gravity and is located in the origin of
our coordinate system. This makes the system non-inertial, the principle upon
which the radial velocity searches for extrasolar planets are based. Although we take the acceleration of the coordinate frame into account, for the low-mass planets the effect is negligible. We take the mass of the star to be 1 \msun.

\subsection{Planet}
Our coordinate frame rotates with the angular velocity of the planet, and 
because we keep the planet on a fixed circular orbit it resides at a fixed
location on the grid, $(r,\theta,\phi)=(r_\mathrm{p},\pi/2,\pi)$. The potential
of the planet is smoothed over 2 grid cells on the highest level of refinement,
which is always much smaller than the Roche lobe of the planet (see Sect. 
\ref{3DsecNum}). This way, we always resolve the potential. The planet is able 
to accrete matter from the disk without changing its dynamical mass. We vary
the mass of the planet between $0.6$ and $300$ \mearths for the isothermal
models, which spans the whole range from the linear regime of Type I 
migration to a gap-opening Jupiter-mass planet. For the 
radiation-hydrodynamical models we focus on a planet of 5 \mearth, which is 
well inside the linear regime \citep[see however][]{2006ApJ...652..730M}.

\subsection{Disk}
The disk is three-dimensional, extending from $r=0.4~r_\mathrm{p}$ to 
$r=2.5~r_\mathrm{p}$ in the radial direction ($r_\mathrm{p}$ is the distance 
from the central star to the planet). In the azimuthal 
direction the computational domain is bounded by $0 \le \phi \le 2\pi$, 
while in the polar direction we go up to $2.5$ pressure scale heights above 
the midplane of the disk: $\pi/2-5H/2r < \theta < \pi/2$. 

\subsubsection{Gas}
The scale height $H$ varies linearly with radius
initially, with $h=H/r=0.05$. The initial temperature profile is found from
assuming hydrostatic equilibrium in the vertical direction and therefore
matches the temperature profile of the isothermal simulations.
The density is also a power law initially, with index $-3/2$. We vary the
midplane density at the location of the planet to investigate different 
cooling regimes. Our nominal value is $10^{-11}$ $\mathrm{g~cm^{-3}}$, 
which is appropriate for the location of Jupiter in the minimum mass solar 
nebula. The initial velocities are zero in the radial and polar 
direction, and the Keplerian speed in the azimuthal direction, with a 
correction for the radial pressure gradient. The equation of state is given 
by the ideal gas law.

We model the disk as an ideal, inviscid fluid, and therefore its evolution
is governed by the Euler equations. These can be cast in the following form:
\begin{equation}
\frac{\partial {\vec W}}{\partial t} + \frac{\partial {\vec F}_r}{\partial r} +
\frac{\partial {\vec F}_\theta}{\partial \theta} + 
\frac{\partial {\vec F}_\phi}{\partial \phi} = {\vec S},
\end{equation}
in which $\vec W$ is the state vector, $\vec F$ are the fluxes in the three
coordinate directions and $\vec S$ is the source vector. The state consists
of the conserved quantities:
\begin{equation}
{\vec W}=r^2 \sin\theta(\rho,~\rho v_r,~\rho v_\theta,~\rho v_\phi,e)^T,
\end{equation}
where $\rho$ is the density, $v_r$ is the radial velocity, $v_\theta$ and
$v_\phi$ are the two angular velocities and $e$ is the total energy density:
\begin{equation}
e=\frac{1}{2}\rho(v_r^2 + r^2 v_\theta^2 + r^2\sin^2\theta (v_\phi+\Omega)^2) -
\rho \Phi + \frac{p}{\Gamma-1}. 
\end{equation}
Here $\Omega$ is the angular velocity of the coordinate frame, $\Phi$ denotes 
the potential, $p$ is the pressure and $\Gamma$ is the adiabatic exponent. 
The three fluxes are given by:
\begin{eqnarray}
{\vec F}_r= r^2 \sin\theta(\rho v_r, \rho v_r^2 + p, \rho v_r v_\theta, 
\rho v_r v_\phi, \rho v_r w)^T\\
{\vec F}_\theta= r^2 \sin\theta(\rho v_\theta, \rho v_r v_\theta, 
\rho v_\theta^2+p/r^2, \rho v_\theta v_\phi, \rho v_\theta w)^T\\
{\vec F}_\phi= r^2 \sin\theta(\rho v_\phi, \rho v_r v_\phi, 
\rho v_\theta v_\phi, \rho v_\phi^2+\frac{p}{r^2\sin^2\theta}, 
\rho v_\phi w)^T,
\end{eqnarray}
where $w$ is the enthalpy of the fluid:
\begin{equation}
w=\frac{e+p}{\rho}.
\end{equation}
The source vector $\vec S$ is then:
\begin{eqnarray}
\label{3DeqSource}
{\vec S}=r^2 \sin\theta \nonumber\\
\left( \begin{array}{c} 0 
\\
\rho r \sin^2 \theta (v_\phi+\Omega)^2 - \rho\frac{\partial{\Phi}}{\partial{r}}
+ \frac{2p}{r} 
\\
-2 \rho \frac{v_r}{r} v_\theta+\sin\theta \cos\theta\rho(v_\phi+\Omega)^2-
\frac{\rho}{r^2}\frac{\partial \Phi}{\partial \theta}+\frac{\cot\theta p}{r^2}
\\
-2 \rho \frac{v_r}{r}(\Omega+v_\phi) - 2\cot\theta\rho v_\theta(\Omega+v_\phi)-
\frac{\rho}{r^2\sin^2\theta}\frac{\partial \Phi}{\partial\phi}
\\
-2\rho(v_r\frac{\partial \Phi}{\partial r} + 
       v_\theta\frac{\partial \Phi}{\partial \theta}+
       v_\phi\frac{\partial \Phi}{\partial \phi})-
\rho\Omega\frac{\partial \Phi}{\partial \phi}
 \end{array}\right).
\end{eqnarray}
The reason for not including an anomalous turbulent viscosity as in 
\cite{2003MNRAS.341..213B} and \cite{2003ApJ...586..540D} is that the
usual $\alpha$-prescription \citep{1973A&A....24..337S} is not
a good description of the magnetic turbulence that is probably the source of 
viscosity \citep{1990BAAS...22.1209B}. The turbulent nature of the disk leads to large temporal and spatial variations in $\alpha$ \citep{2003MNRAS.339..983P}. Neglecting viscosity altogether means
that our results are valid in regions of the disk where the ionization 
fraction is not high enough to sustain magnetic turbulence.

\subsubsection{Radiation}
Planets that do not open gaps ($M_\mathrm{p}\ll $ \mjup) are deeply embedded in 
the disk where the material is very optically thick. It is therefore 
appropriate to consider radiative transfer in the diffusion limit. However, 
because the density drops more than an order of magnitude over a few pressure 
scale heights above the midplane it is necessary to deal with the optically 
thin regime as well. 

We start by considering a medium at rest, i.e. we set the velocity of the 
gas to zero. In the diffusion limit there is essentially no angular information
in the radiation field, which is then governed by a diffusion equation for its 
energy density $E$:
\begin{equation}
\label{3DeqDiff}
\frac{\partial E}{\partial t} - \nabla \cdot \frac{c}{3\kappa} \nabla E =
\kappa c (B-E),
\end{equation}
where $c$ denotes the speed of light, $\kappa$ is the opacity in 
$\mathrm{cm^{-1}}$ and $B$ is the radiative source term. We consider only 
thermal emission from the disk, therefore
\begin{equation}
B=a~T^4,
\end{equation}
where $a$ is the radiation constant and $T$ is the gas temperature. In order 
to conserve total energy an extra source term $S_\mathrm{rad}=-\kappa c (B-E)$
appears in the gas energy equation. We work with a Rosseland mean opacity that 
is typical for the MMSN \citep[see][]{1994ApJ...427..987B}. Initially we set
$E=B$ in all simulations.

In the optically thin streaming limit the evolution of the radiation energy
is given by:
\begin{equation}
\label{3DeqStream}
\frac{\partial E}{\partial t} + \nabla \cdot (c~E {\vec{\hat n}}) =
\kappa c (B-E),
\end{equation}
where $\vec{\hat n}$ is a unit vector in the direction of propagation of the
light front. Note first of all that this is an advection equation rather than
a diffusion equation, and second that there is a strong angular dependence
of the radiation field. Naively applying Eq. \eqref{3DeqDiff} in the optically
thin case will lead to large errors.

However, because the most important part of the disk is optically thick it
is still useful to start with Eq. \eqref{3DeqDiff} but to avoid divergences in 
the optically thin case. Note that avoiding divergences is different from 
obtaining the correct solution, which would require a method based on discrete 
ordinates or on a Monte Carlo algorithm. Both of these methods have great 
difficulties with the case of a very optically thick disk.

We write Eqs. \eqref{3DeqDiff} and \eqref{3DeqStream} in the more general form:
\begin{equation}
\label{3DeqStatic}
\frac{\partial E}{\partial t} + \nabla \cdot {\vec F}_\mathrm{R} =
\kappa c (B-E),
\end{equation}
in which the radiative flux ${\vec F}_\mathrm{R}$ is given by
\begin{equation}
\label{3DeqFlux}
{\vec F}_\mathrm{R}=-\frac{c~\lambda}{\kappa}\nabla E,
\end{equation}
where $\lambda$ is called the flux limiter. The definition of 
${\vec F}_\mathrm{R}$ in Eq. \eqref{3DeqFlux} in terms of $E$ removes the need 
for solving a differential equation for ${\vec F}_\mathrm{R}$. Closing the radiative 
moment equations this way is called flux-limited diffusion (FLD)
\citep{1981ApJ...248..321L}. The flux-limiter $\lambda$ deals with the 
transition from the diffusion limit to the streaming limit. The exact 
functional form of $\lambda$ also implicitly determines the angular dependence 
of the radiation field. For an overview of different forms see 
\cite{1989A&A...208...98K}. We have used
\begin{equation}
\lambda=\left\{ \begin{array}{ll}
       \frac{2}{3+\sqrt{9+10~R^2}}& \mbox{for~ $0\leq R \leq 2$};\\
       \frac{10}{10~R+9+\sqrt{180~R+81}} & \mbox{for~ $R>2$},
\end{array} \right. 
\end{equation}
where
\begin{equation}
R=\frac{1}{\kappa}\frac{\left|\nabla E\right|}{E},
\end{equation}
but the results do not depend sensitively on the adopted form of $\lambda$,
mainly because of the high optical depth of the disk near the planet.

Flux-limited diffusion typically gives poor results in the streaming limit 
\citep{2003ApJS..147..197H}, but for our case of a deeply embedded planet
it is a reliable and relatively cheap method for doing radiative transfer.

When the velocity of the gas is taken into account Eq. \eqref{3DeqStatic} needs
to be extended in two ways to account for advection of radiative energy
and momentum transfer between gas and radiation. The full equation for 
the evolution of the radiative energy then reads:
\begin{equation}
\label{3DeqRadEnergy}
\frac{\partial E}{\partial t} + \nabla \cdot ({\vec F}_\mathrm{R}+
{\vec v}~E+{\vec v}
\cdot \mathcal{P}) = \kappa c (B-E) -\frac{\kappa}{c} {\vec v} \cdot 
{\vec F}_\mathrm{R},
\end{equation}
where $\mathcal{P}$ is the radiative stress tensor, given by
\begin{equation}
\mathcal{P}=E \mathcal{T}_\mathrm{Edd}.
\end{equation}
The Eddington tensor $\mathcal{T}_\mathrm{Edd}$ is calculated using
\begin{equation}
\mathcal{T}_\mathrm{Edd}=\frac{1}{2}\left[ (1-f_\mathrm{Edd})\mathcal{I}
+(3~f_\mathrm{Edd}-1){\vec{\hat n}}{\vec{\hat n}} \right],
\end{equation}
where ${\vec{\hat n}}=\nabla E/|\nabla E|$. The Eddington factor 
$f_\mathrm{Edd}$ can be calculated from the streaming factor $f_\mathrm{s}=
|{\vec F}_\mathrm{R}|/cE$:
\begin{equation}
f_\mathrm{Edd}=\left\{ \begin{array}{ll}
       \frac{1}{3}+\frac{1}{6} f_\mathrm{s}^2 & 
       \mbox{for~ $0\leq f_\mathrm{s} \leq 0.4$};\\
       \frac{5}{9}(1-f_\mathrm{s})^2+ f_\mathrm{s}^2 & 
       \mbox{for~ $0.4\leq f_\mathrm{s} \leq 1$}. 
\end{array} \right.
\end{equation}

The extra source vector in the gas flow equations ${\vec S}_\mathrm{rad}$
is given by:
\begin{eqnarray}
\label{3DeqSourcerad}
{\vec S}_\mathrm{rad}=r^2 \sin\theta
\left( \begin{array}{c} 0 \\
\frac{\kappa}{c} F_\mathrm{R,r} \\
\frac{\kappa}{c} F_\mathrm{R,\theta} \\
\frac{\kappa}{c} F_\mathrm{R,\phi} \\
-\kappa~c(B-E)
\end{array}\right).
\end{eqnarray}
The total source term for the gas is given by $\vec{S}+\vec{S}_\mathrm{rad}$.

\section{Numerical method}
\label{3DsecNum}
We use the technique of operator splitting to evolve the hydrodynamic part
and the radiative part separately every time step. 

\subsection{Hydrodynamics}
We numerically solve the flow equations for the gas using a three-dimensional
version of the RODEO method \citep{2006A&A...450.1203P}. For isothermal 
simulations the necessary eigenvalues, eigenvectors and projection coefficients
are simple generalizations of the two-dimensional analogs given in 
\cite{2006A&A...450.1203P}. For simulations including an energy equation, the 
method is equivalent to the non-relativistic version of the general 
relativistic method of \cite{1995A&AS..110..587E}. 

In short, RODEO evolves the Euler equations using an approximate Riemann
solver \citep[the Roe solver, see][]{1981...............}, together with
stationary extrapolation to account for the geometrical and hydrodynamical
source terms (Eq. \eqref{3DeqSource}). The Coriolis forces are treated in an
exact fashion to enforce exact angular momentum conservation \citep{1998A&A...338L..37K,
2006A&A...450.1203P}. The extra source terms due to interaction with the 
radiation field (Eq. \eqref{3DeqSourcerad}) are treated as external source 
terms and are integrated separately (see below). 

A module for adaptive mesh refinement (AMR) is used to obtain high resolution close to the planet. Because the planet remains at a fixed location on the grid, this treatment amounts to having a nested grid.

Accretion onto the planet is handled in the same way as in 
\cite{2003ApJ...586..540D}: every time step $\Delta t$ the density near the
planet is reduced by a factor $1-f\Delta t$. The area from which we take 
away mass has a size $r_\mathrm{acc}$ times the Hill radius of the planet. For all accretion results in this paper, we have used $f=5/3$ and  $r_\mathrm{acc}=0.1$ 
\citep[see][]{2002A&A...385..647D}. \cite{2006A&A...445..747K} used, in 
addition to runs with this accretion prescription, a procedure that conserves 
total energy locally by adding the kinetic and potential energy that is 
accreted onto the planet to the internal energy of the gas. Because we solve 
for the total energy of the gas directly, our accretion prescription does 
conserve total energy. This means that in the accretion region the gas is 
heated by the accretion process. 

When calculating the torque on the planet, we exclude material that resides
within the Hill sphere of the planet. This was also done in 
\cite{2003MNRAS.341..213B}. Although material inside the Hill sphere may
exert a torque on the planet, which may lead to migration reversal 
\citep{2003ApJ...586..540D}, this torque is very sensitive to numerical
resolution and the details of the accretion process. Moreover, this is also
the region where self-gravity should play an important role, which is
not included in our models. Therefore we focus on torques due to material
\emph{outside} the Hill sphere of the planet. We anticipate that for the
low-mass planets that we consider in this paper this will not affect our
results.
 
\subsection{Flux-limited diffusion}
Because of the intrinsic short time scale for radiative effects, which is due
to the enormously fast communication speed $c$ compared to the sound speed,
it is necessary to integrate the equation for $E$ in an implicit way. 

First of all, consider the gas coupling term. We follow the approach described
in \cite{2003ApJS..147..197H}. The gas is heated by the radiation field
according to
\begin{equation}
\frac{\partial \epsilon}{\partial t}=\kappa c (E-B),
\end{equation}
where $\epsilon=p/(\Gamma-1)$ is the internal energy of the gas. Upon 
differencing  both sides we obtain:
\begin{equation}
\label{3DeqGasEnergy}
\epsilon^{n+1}-\epsilon^n=\Delta t \kappa c (E^{n+1}-B^{n+1}),
\end{equation}
where $X^n$ denotes quantity $X$ evaluated at time n. We approximate
$B^{n+1}$ using a Taylor series expansion:
\begin{equation}
\label{3DeqTaylorB}
B^{n+1} \approx a(T^n)^3~(4T^{n+1}-3T^n),
\end{equation}
and we spell out the internal energy as
\begin{equation} 
\label{3DeqInt}
\epsilon^{n+1}=\frac{p^{n+1}}{\Gamma-1}=
\frac{\rho \mathcal{R} T^{n+1}}{\Gamma-1},
\end{equation}
in which $\mathcal{R}$ is the universal gas constant and $\Gamma$ is the 
adiabatic exponent.

Putting Eq. \eqref{3DeqInt} and \eqref{3DeqTaylorB} into Eq. 
\eqref{3DeqGasEnergy} we can solve for $T^{n+1}$:
\begin{equation}
T^{n+1}=\frac{\epsilon^{n} + \Delta t \kappa c E^{n+1}
+ 3 \Delta t \kappa c B^{n}}{\epsilon^{n}+ 
4\Delta t \kappa c B^{n}}~T^{n}.
\end{equation}
This expression is used in Eq. \eqref{3DeqTaylorB} to evaluate 
$B^{n+1}$. We are now in a position to difference Eq. 
\eqref{3DeqRadEnergy} \citep[for details see][]{2003ApJS..147..197H}, which 
gives us an algebraic equation for $E^{n+1}_{i,j,k}$ at grid 
position $(i, j, k)$:
\begin{eqnarray}
\label{3DeqRadFinite}
\begin{array}{l}
d_{i,j,k}   E^{n+1}_{i,j,k} +
d_{i+1,j,k} E^{n+1}_{i+1,j,k} +
d_{i,j+1,k} E^{n+1}_{i,j+1,k} + \\
d_{i,j,k+1} E^{n+1}_{i,j,k+1} +
d_{i-1,j,k} E^{n+1}_{i-1,j,k} +
d_{i,j-1,k} E^{n+1}_{i,j-1,k} + \\
d_{i,j,k-1} E^{n+1}_{i,j,k-1} =
Q^{n}_{i,j,k},\\ 
\end{array}
\end{eqnarray}
for certain $d_{i,j,k}$. Note that we use only a seven point
finite difference stencil. This is cheaper than taking into account all
26 neighboring cells, while we found no noticeable difference in the
result. 

We can write Eq. \eqref{3DeqRadFinite} as a linear system equation:
\begin{equation}
\label{3DeqLinear}
\mathcal{D}~{\vec E}^{n+1}={\vec Q}^{n},
\end{equation}
in which $\mathcal{D}$ is a $N\times N$ matrix and ${\vec E}^{n+1}$ 
and ${\vec Q}^{n}$ are vectors with $N$ elements. Here, $N$ is the 
total number of grid points. Fortunately, this huge matrix $\mathcal{D}$ is 
sparse; it has only seven non-zero entries every row. For such matrices 
efficient iterative linear solvers have been developed in the literature.
For an overview see \cite{barrett}.

The most basic characteristic of a sparse linear solver is whether it is a
\emph{stationary} method or not. Stationary methods for solving Eq. 
\eqref{3DeqLinear} can be cast in the form:
\begin{equation}
\label{3DeqStatUpdate}
{\vec E}^{n+1}_m = \mathcal{A} {\vec E}^{n+1}_{m-1}+{\vec C},
\end{equation}
where ${\vec E}^{n+1}_m$ is the approximation to ${\vec E}^{n+1}$
after $m$ iterations. The matrix $\mathcal{A}$ nor the vector $\vec C$ depends 
on $m$ for a stationary method. Examples are the Jacobi method, the 
Gauss-Seidel method and successive over-relaxation (SOR), which was used for 
example in \cite{2006A&A...445..747K}. The main disadvantage in using SOR lies 
in the fact that the convergence behavior of the method depends on a 
relaxation parameter $\omega$, for which the optimal value is 
problem-dependent.

Non-stationary methods are a more recent development, and they differ from
stationary methods in that the structure of the update step changes with 
each iteration. A simple case is given by Eq. \eqref{3DeqStatUpdate} with a 
matrix $\mathcal{A}$ and vector $\vec C$ that do depend on $m$. Examples of 
non-stationary methods are the conjugate gradient method (CG), the generalized 
minimal residual method (GMRES), the quasi minimal residual method (QMR) and
the bi-conjugate gradient stabilized method (Bi-CGStab).

The optimal method for a given problem depends on the structure of the matrix
$\mathcal{D}$. CG, for example, is only applicable to symmetric positive
definite matrices, while we know in advance that due to opacity gradients the 
matrix $\mathcal{D}$ will not be symmetric. One could try CG on the matrix
$\mathcal{D}^T \mathcal{D}$, where $\mathcal{D}^T$ is the transpose of 
$\mathcal{D}$, but this is not a good idea because the condition number of
$\mathcal{D}^T \mathcal{D}$ is the square of the condition number of
$\mathcal{D}$, which would leads to very slow convergence of the method.

Another difference amongst the various methods is the memory requirement.
When dealing with large matrices this can be a real bottleneck. For GMRES
it can be shown that the optimal solution is reached within a finite number
of iterations, but these iterations become more and more expensive because
the method needs information from \emph{all} previous steps. To avoid the
enormous storage requirements one could decide to restart GMRES after a 
fixed number of iterations, but this introduces a free parameter in the 
method that needs to be tuned to the problem. Finally, QMR is more expensive 
in memory than Bi-CGStab in terms of memory because it needs more auxiliary
variables. Therefore we have chosen Bi-CGStab as our method.

One potential disadvantage of Bi-CGStab is that it needs four inner products
every iteration. An inner product is a global quantity, and therefore it
requires communication between different processors, in addition to the 
boundary conditions. They act as synchronization points, and therefore 
load-balancing is critical in this part of the numerical method. We have
implemented a dynamical load-balancing module, in which during execution 
of the code the processors exchange part of the computational domain to 
optimize the domain decomposition. Note that due to the AMR it is not true
that every processor should get an equal part of the computational domain,
because at an AMR boundary more work is needed (i.e. interpolation of 
data, flux correction). 

\begin{figure}
%\resizebox{\hsize}{!}{\includegraphics[bb=255 10 495 241]{paardekooperfig1.eps}}
\resizebox{\hsize}{!}{\includegraphics[]{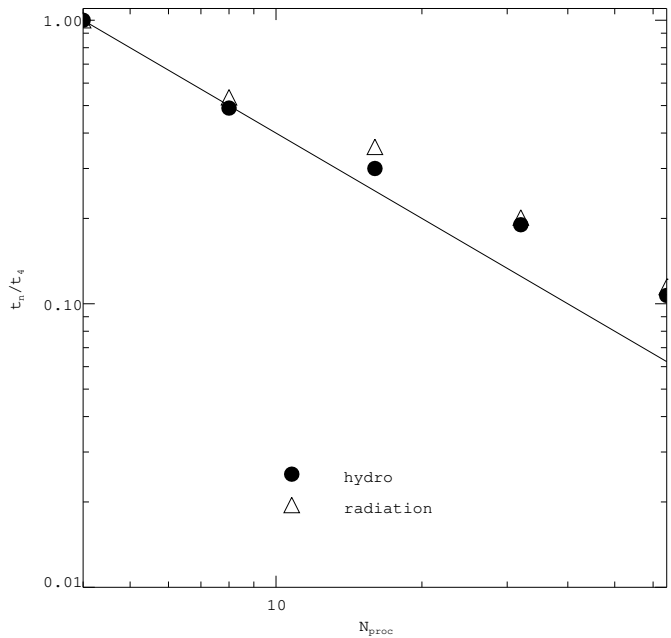}}
\caption{Strong scaling test for the three-dimensional radiation-hydrodynamical
code, shown as the total execution time (relative to the execution time using
four processors) as a function of the total number of processors. The solid 
line indicates perfect scaling, i.e. running on twice as many
processors results in half the original execution time. Filled circles indicate
the hydrodynamics part of the code; open triangles the radiation part. Note 
that both parts of the code scale equally well up to 64 processors.}
\label{3Dfig1}
\end{figure}

We have preformed a strong scaling test on a low-resolution disk with a
low-mass planet. The main grid consisted of 128 radial cells, 384 azimuthal 
cells and 8 meridional cells, on top of which we put 2 levels of AMR. The
simulation was run on different numbers of processors ($n_\mathrm{proc}$) for 
a total of 5 orbits of the planet, which corresponds to approximately 7500 time 
steps. The resulting total execution times, relative to the case of 4 
processors, is shown in Fig. \ref{3Dfig1}. Focusing on the hydrodynamical part 
first (filled circles), we see that the code scales perfectly up to 
$n_\mathrm{proc}=8$. Due to memory requirements the simulation could not run 
on less than 4 processors, but the perfect scaling from 4 to 8 processors 
indicates that the code should scale very well from 1 to 4 processors as well. 

\begin{table*}
\caption{Overview of model parameters. From left to right, the columns 
indicate: planet mass in \mearth, value of the adiabatic exponent ($\Gamma=1$
means a locally isothermal equation of state), viscosity parameter $\alpha$,
radiation flag, accretion flag and the number of refinement levels. When
$n_\mathrm{lev}$ levels of refinement are used, the local resolution within
a distance of $2H$ from the planet is a factor $2^{n_\mathrm{lev}}$ higher
than the base resolution of $\Delta r=0.0082~r_\mathrm{p}$. For the local 
models, the indicated resolution is valid for the whole computational domain.}
\label{table:1}
\centering
\begin{tabular}{c c c c c c c c}
\hline\hline
$M_\mathrm{p}$ (\mearth) & $\Gamma$ & $\alpha$ & FLD & Acc & $n_\mathrm{lev}$
& $\mathcal{S}$ & Comments \\
\hline
0.6 & 1 & 0.004 & n & y & 5 & 0 &\\
5   & 1 & 0.004 & n & y & 4 & 0 &\\
5   & 1 & 0.004 & n & y & 3 & 0 &\\
5   & 1 & 0.004 & n & y & 2 & 0 &\\
40  & 1 & 0.004 & n & y & 3 & 0 &\\
320 & 1 & 0.004 & n & y & 2 & 0 &\\
\hline
5   & 1.005 & 0 & n & y & 4 & -0.9925 &\\ 
5   & 1.01  & 0 & n & y & 4 & -0.9850 &\\ 
5   & 1.005 & 0 & n & y & 3 & -0.9925 &\\ 
5   & 1.01  & 0 & n & y & 3 & -0.9850 &\\ 
\hline
0.6 & 1.4   & 0 & y & n & 5 & -0.4 &\\
0.6 & 1.4   & 0 & y & y & 5 & -0.4 &\\
0.6 & 1.4   & 0 & y & n & 4 & -0.4 &\\
0.6 & 1.4   & 0 & y & y & 4 & -0.4 &\\
5   & 1.4   & 0 & y & n & 4 & -0.4 &\\
5   & 1.4   & 0 & y & y & 4 & -0.4 &\\
5   & 1.4   & 0 & y & n & 3 & -0.4 &\\
5   & 1.4   & 0 & y & y & 3 & -0.4 &\\
5   & 1.4   & 0 & y & n & 2 & -0.4 &\\
5   & 1.4   & 0 & y & y & 2 & -0.4 &\\
5   & 1.4   & 0 & y & n & 4 & -0.4 & 10 times lower density\\
5   & 1.4   & 0 & y & n & 4 & -0.4 & 100 times lower density\\
\hline
5   & 1.4   & 0 & n & n & 4 & -0.4 &local model\\
5   & 1     & 0 & n & n & 4 & 0 & local model\\
5   & 1.4   & 0 & n & n & 4 & -0.2 & local model\\ 
5   & 1.4   & 0 & n & n & 4 & 0.2 & local model\\ 
5   & 1.4   & 0 & n & n & 4 & 0.4 & local model 
\end{tabular}
\end{table*}

When we increase $n_\mathrm{proc}$ beyond 8, the scaling is still
very good, although not perfect. This is always the case at some point in
a strong scaling test, because the amount of data that needs calculation stays
the same, while the amount of data that has to be communicated increases
with the number of processors. At some point increasing $n_\mathrm{proc}$
does not result in a reduction in execution time anymore. Fortunately, for
up to 64 processors we still measure a significant speed-up. This gives us 
confidence that for the high-resolution runs (that have 8 times more 
computational cells in the main grid) the scaling will be good up to 
$8 \times 64 = 512$ processors. The simulations presented in this paper were 
run on 126 processors of an SGI Altix 3700 CC-NUMA system with Intel Itanium-2 processors of $1.3$ GHz. No use was made of the shared memory capacities of this system.

An interesting point in Fig. \ref{3Dfig1} is that the radiation module scales 
equally well as the hydrodynamics, even though the numerical methods for both 
parts of the physical problem differ significantly. This is an indication 
that for this problem only a small number of iterations is needed in the 
linear solver to reach convergence.

\subsection{Boundary conditions}
The planet excites waves in the disk, which subsequently propagate to the
boundaries of the computational domain. To minimize reflection of outgoing 
waves we employ the non-reflective boundary conditions as outlined in
\cite{2006A&A...450.1203P} for the hydrodynamic variables. At the midplane 
we use reflective boundary conditions. No reflected waves were observed near 
the radial and upper meridional ($\theta=\pi/2-5h/2$) boundaries.

For the boundary conditions for the radiation energy density we take a
different approach, because the initial condition for the gas temperature
does not correspond to a stationary solution for $E$. In fact, due to the
different opacity regimes at different heights above the midplane of the
disk it is difficult, if not impossible, to obtain a stationary solution for 
$E$. In the case where the disk is optically thin throughout, the radiation
energy density should vary as $E \propto r^{-2}$, which means that matter
in equilibrium with this radiation has a temperature that varies as 
$T \propto r^{-1/2}$. In vertical hydrostatic equilibrium the relative scale
height $h$ is then proportional to $r^{1/4}$, while the initial condition
for the gas is a constant $h$. However, all disk models quickly (within 
a few orbits of the planet) adapt to a new equilibrium corresponding to the
temperature structure set by the radiation field. 

\begin{figure*}
\centering
\includegraphics[width=17cm]{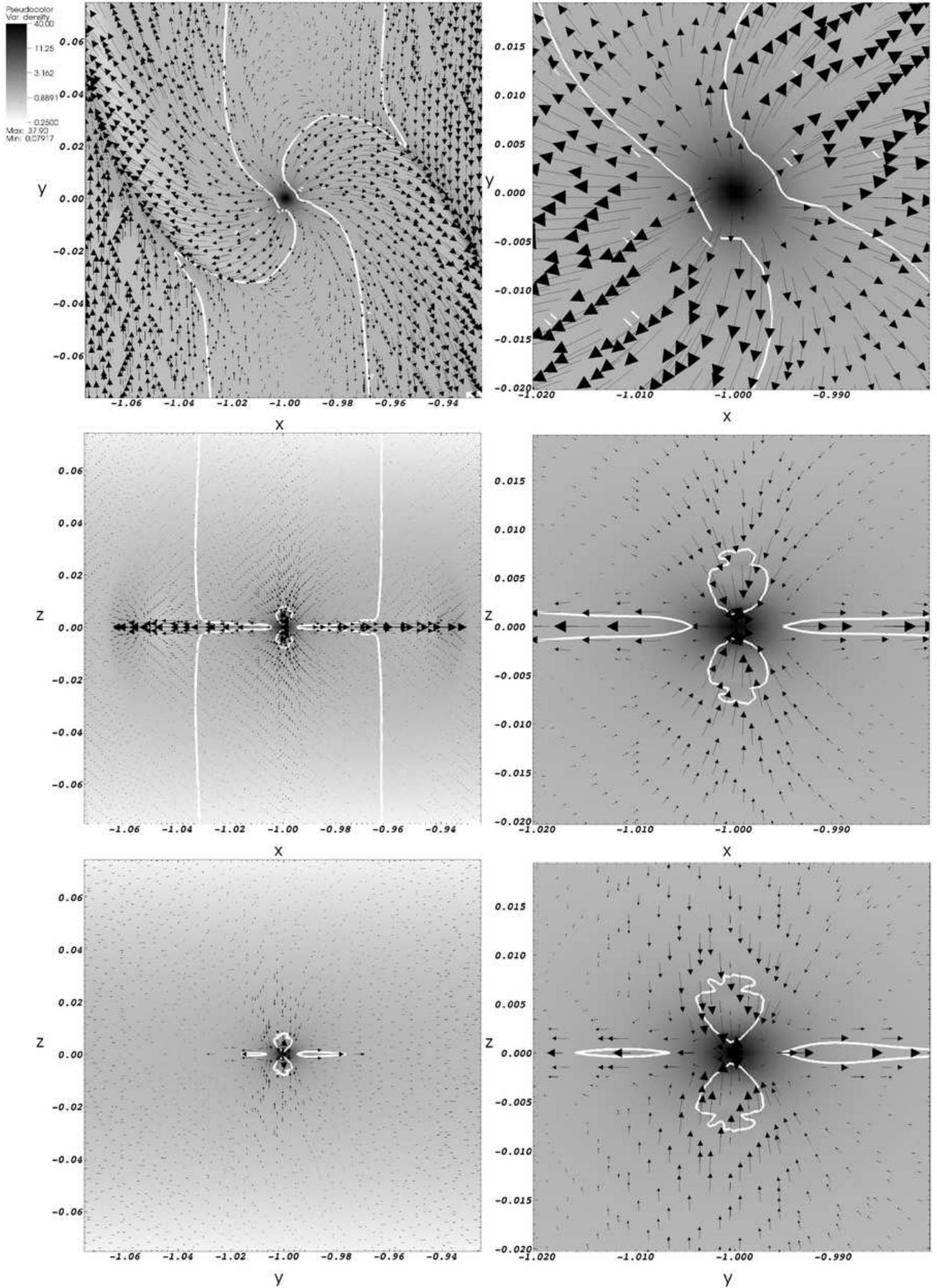}
\caption{Density and velocity fields around an accreting 5 \mearths planet
after 10 orbits. Top panels: slice through $z=0$ ($\theta=\pi/2$), middle 
panels: slice through $y=0$ ($\phi=\pi$), bottom panels: slice through $x=-1$
($r=1$). Inside the white contour the velocities are supersonic. }
\label{3Dfig2}
\end{figure*}

Real protoplanetary disks are heated internally by viscous dissipation
(in regions where magnetic turbulence operates) and externally by radiation
from the central star. Irradiation effects may play a large role in structuring
protoplanetary disks \citep{2000A&A...361L..17D,2001ApJ...560..957D}, but a
detailed treatment of these effects is beyond the scope of this paper. 
However, in absence of heating sources the disk cools down rapidly 
\citep[see][]{2006A&A...445..747K}, which will change the torque on embedded
planets. 

To account for irradiation in an approximate way we set the radiation energy 
density at the inner boundary to a constant value, which, for an optically
this disk, would correspond to $h=0.05$ at the location of the planet. This
way, we can compare the radiation-hydrodynamical simulations directly to 
their isothermal counterparts.

At the midplane we use reflective boundary conditions, while at the outer
radial boundary we set the ghost cells such that $F_{\mathrm{R},r} \propto
r^2$, which corresponds to an outflow boundary. At the $\theta=\pi/2-5h/2$
boundary we apply a similar procedure. This way, radiation can propagate 
freely off the grid.

\subsection{Test problems}
The RODEO hydrodynamics module was tested extensively on a Cartesian grid
with shock tube problems: in all coordinate directions separately as well as
in two dimensions diagonally. Also a genuinely two-dimensional problem, the
wind tunnel with step, was studied and the results showed that the internal
Roe solver is able to deal with the complex shock structures that arise in
this problem \citep[see also][]{1991A&A...252..718M}.
On the two-dimensional disk-planet problem, RODEO was compared to
other methods in \cite{comparison}.

The radiation module was tested with one- and two-dimensional diffusion
problems with a simplified opacity law ($\kappa=$ constant). See 
\cite{1992ApJS...80..819S} for a description of the diffusion tests. The
numerical results were indistinguishable from the analytical solution.

Another test for the radiation solver is an optically thin circumstellar
disk. Although FLD is in principle not very well suited for optically thin
problems, it should nevertheless let $E$ diffuse towards the correct solution
$E \propto r^{-2}$. Starting from the initial condition $E \propto r^{-4}$
(for $h=$ constant, $T \propto r^{-1}$, therefore $E = a T^4 \propto r^{-4}$)
the numerical solution agrees with $E \propto r^{-2}$ within one orbit of
the planet after which it remains stable. This is an important test, because 
it shows that we can compare the radiation-hydrodynamical results in the
optically thin limit directly to the isothermal runs.

Finally, we considered an optically thick disk with a constant opacity. In 
this case, the equilibrium solution is given by $E \propto r^{-1}$. Again,
starting from the initial condition for constant $h$ the numerical solution
agrees with  $E \propto r^{-1}$ within a few orbits of the planet, depending
on the exact value of the opacity. 

\section{Results}
\label{3DsecRes}
In the previous sections we have described the physical and numerical setup
for the three-dimensional radiation-hydrodynamical simulations. However, 
we now move from the simple, 2D, viscous simulations toward the full models in
two steps. The first step consists of 3D, isothermal simulations that are
compared to \cite{2003MNRAS.341..213B} and \cite{2003ApJ...586..540D}. After
that, we will include the energy equation but treat the cooling in a very
simplistic way by setting the adiabatic exponent $\Gamma$ close to unity.
Finally, we will present the full radiation-hydrodynamical results.
In Table \ref{table:1} we give an overview of the simulations
discussed in this paper.

\begin{figure}
%\resizebox{\hsize}{!}{\includegraphics[bb=24 24 595 570]{paardekooperfig3.eps}}
\resizebox{\hsize}{!}{\includegraphics[]{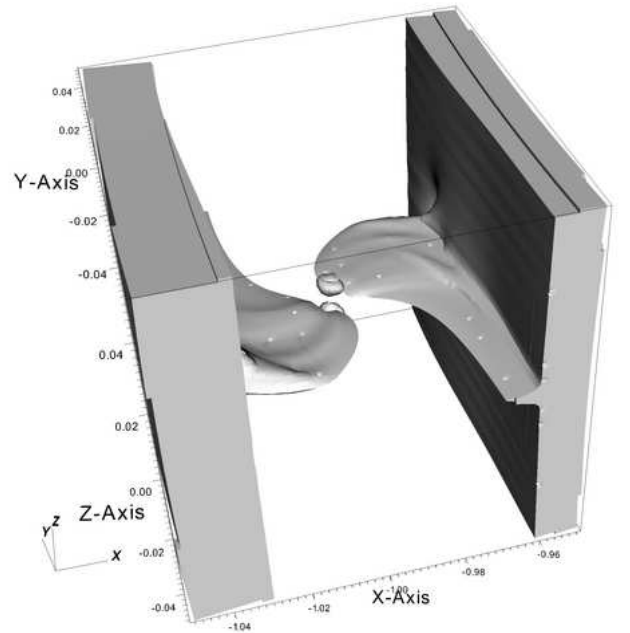}}
\caption{Isovolume for a Mach number of 1. Inside the solid structures the 
velocity relative to the planet is supersonic. }
\label{3Dfig3}
\end{figure}

\subsection{Isothermal models}
\label{3DsecIso}
The flow structure around planets in three-dimensional isothermal disks has
been described before \citep{2003MNRAS.341..213B,2003ApJ...586..540D}. 
Therefore we limit the discussion to a few interesting remarks on the 
3D-structure of the accretion flow. The disk model we use differs from the one
described in Sect. \ref{3DsecModel} in two ways: we use a locally isothermal
equation of state, and a kinematic $\alpha$-type viscosity 
\citep{1973A&A....24..337S} with $\alpha=0.004$ at the location of the planet. 
Essentially this model is the same as in \cite{2003ApJ...586..540D}.

In Fig. \ref{3Dfig2} we show density slices along the three coordinate axis for
a 5 \mearths planet after 10 orbital periods. By this time the flow has reached
a steady state \citep{2003MNRAS.341..213B}. For a low-mass planet like this
the spiral wave features in the density are not strong, but they appear clearly
in the velocity field. 

From the middle and bottom panels of Fig. \ref{3Dfig2} we see that the material
accreting onto the planet originates predominantly above and below the planet. 
As the material enters the Hill sphere of the planet, which for this mass is
$0.017~r_\mathrm{p}$ large, it is accelerated to supersonic velocities. This
was also observed by \cite{2003ApJ...586..540D}, albeit not for a planetary
mass as low as 5 \mearth. Interesting is the equatorial outflow of mass that is
apparent in Fig. \ref{3Dfig2}. As the accreting material rains down on the 
planet it squeezes the envelope which forces some material to be expelled
from the envelope in the equatorial plane of the disk. Even before leaving the
Hill sphere of the planet this material again reaches supersonic velocities.
It is important to note that no vertical hydrostatic equilibrium is established
near the planet. This was also observed by \cite{2003ApJ...586..540D}, even for
non-accreting planets. The appearance of outflows does not depend on the 
boundary at $\theta=\pi/2$: we found the same behavior for a disk extending
from $\theta=\pi/2-5h/2$ to $\theta=\pi/2+5h/2$. Therefore, this outflow is not linked to the boundary conditions.

\begin{figure}
%\resizebox{\hsize}{!}{\includegraphics[bb=30 60 570 575]{paardekooperfig4.eps}}
\resizebox{\hsize}{!}{\includegraphics[]{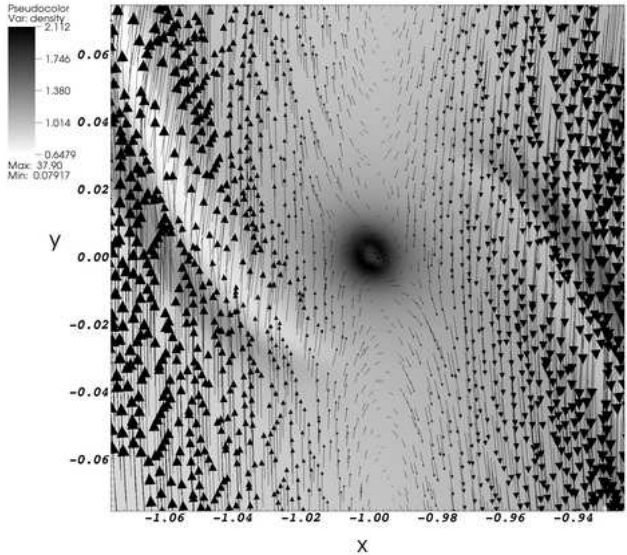}}
\caption{Density and velocity fields around an accreting 5 \mearths planet
after 10 orbits. The slice is taken at a distance of $0.005~r_\mathrm{p}$ 
from the mid plane of the disk.}
\label{3Dfig4}
\end{figure}

\begin{figure}
%\resizebox{\hsize}{!}{\includegraphics[bb=30 60 570 575]{paardekooperfig5.eps}}
\resizebox{\hsize}{!}{\includegraphics[]{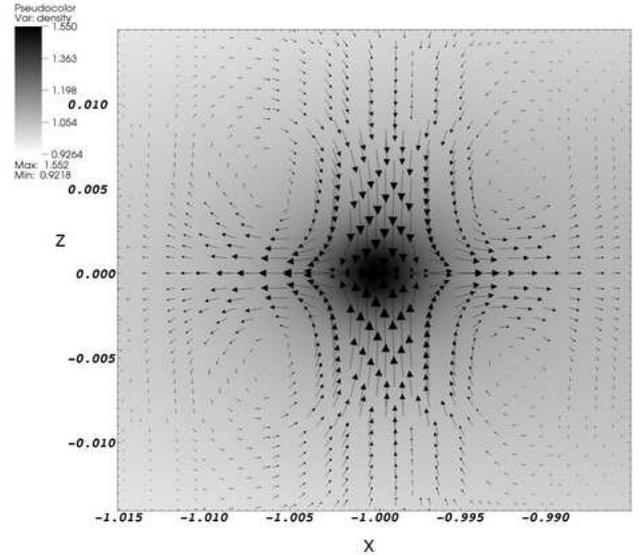}}
\caption{Slice through $\phi=\pi$ for a $0.6$ \mearths planet after 10 orbits,
showing density and velocity. For this low-mass planet all velocities are
subsonic.}
\label{3Dfig5}
\end{figure}

In Fig. \ref{3Dfig3} we show the three-dimensional structure of the supersonic
accretion and subsequent equatorial outflow. The solid volume represents the
region where the Mach number of the flow, relative to the planet, is larger
than one. For large radial distances to the planet the flow is always 
supersonic due to the Keplerian shear; this leads to the two large pizza boxes
on both radial sides of Fig. \ref{3Dfig3}. The supersonic accretion velocities
appear as the two small spheres just above and below the planet. The 
equatorial outflow is quickly turned into spiral wave features by the 
Keplerian shear. Note that the outflow is confined to the midplane of the
disk. This is illustrated further in Fig. \ref{3Dfig4}, where we show a density
slice parallel to the midplane but at a height of $0.005~r_\mathrm{p}$. Note
that this is well within the Hill sphere of the planet. Even so, the velocity
structure differs dramatically from that in the midplane. Apart from the
unperturbed Keplerian velocity on both sides of the planet we can also see
the horse-shoe orbits, as well as a rotating accretion flow close to the 
planet. This was also found by \cite{2003ApJ...586..540D}. 

The velocity structure inside the Hill sphere of the planet is intrinsically
linked to the accretion procedure \citep{2003ApJ...586..540D}. However, we
found that this is not true for the equatorial outflow. It is there always 
when the vertical accretion flow is, albeit with a different magnitude for
different accretion prescriptions. 

\begin{figure*}
\includegraphics[width=17cm]{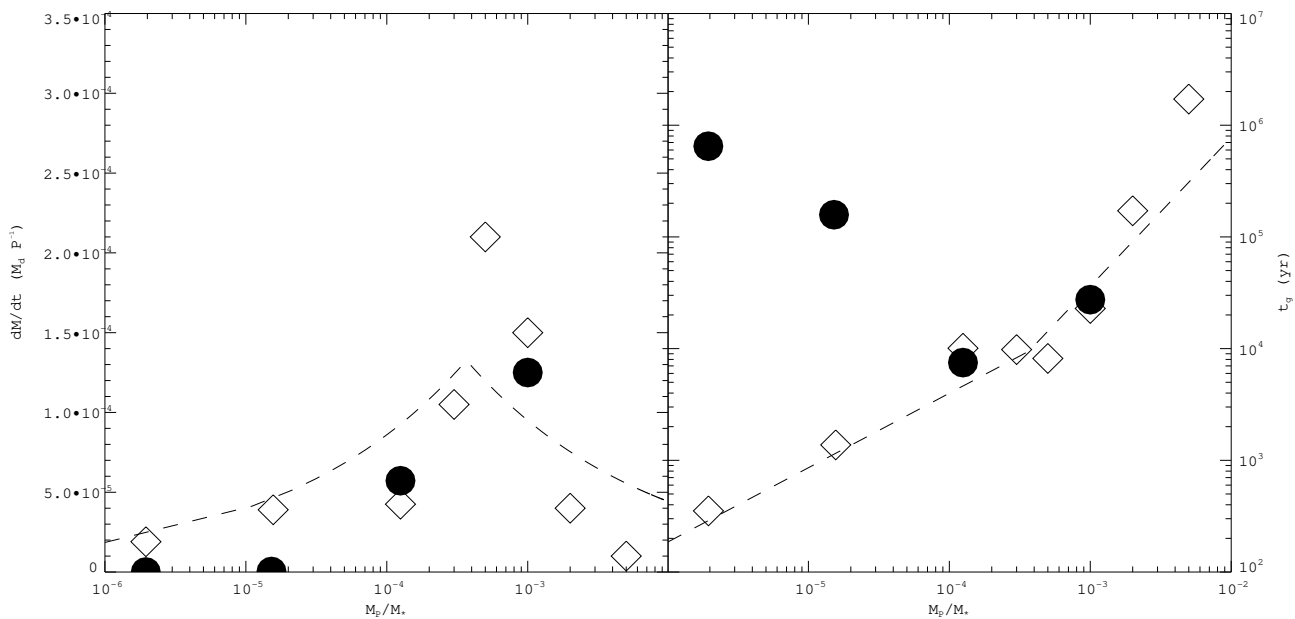}
\caption{Accretion rates for two- and three-dimensional isothermal 
simulations. Open symbols represent the two-dimensional results from 
\cite{2006A&A...450.1203P}, filled symbols denote the three-dimensional 
results. Left panel: accretion rate in units of disk masses per orbit for 
different planetary masses. Right panel: growth time scales $\tau_\mathrm{g}=
M_\mathrm{p}/\dot M_\mathrm{p}$ in years, where we assume a disk mass of 5 
\mjup.}
\label{3Dfig6}
\end{figure*}

For a planet of $0.6$ \mearths the flow within the Hill sphere remains 
subsonic. The density and velocity field near the planet is shown in Fig.
\ref{3Dfig5}. Again we can see the vertical accretion flow from above and below
the planet, some of which is deflected into the midplane of the disk where
it may leave the Hill sphere of the planet. We can identify four circulation
patterns near the edge of the Hill sphere, which is approximately $0.009~
r_\mathrm{p}$ large. Some of these patterns can also be seen in the results 
of \cite{2003ApJ...586..540D}.

\begin{figure}
%\resizebox{\hsize}{!}{\includegraphics[bb=257 10 500 241]{paardekooperfig7.eps}}
\resizebox{\hsize}{!}{\includegraphics[]{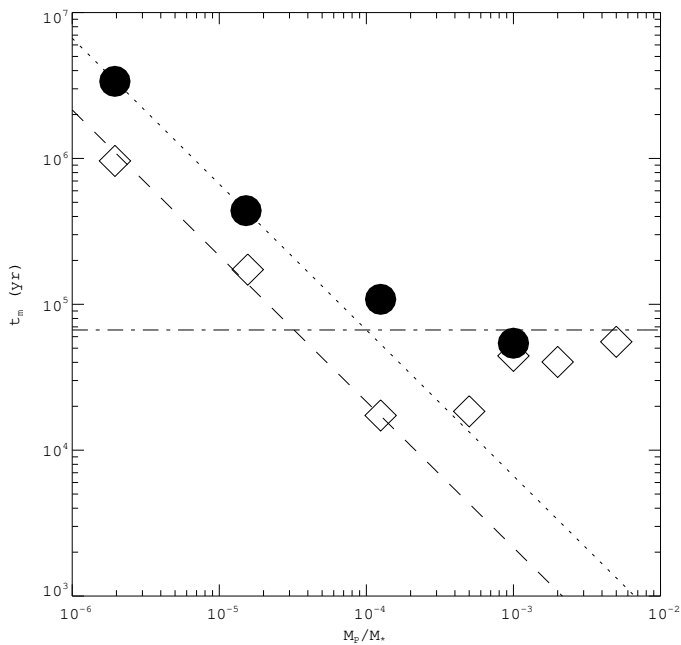}}
\caption{Migration time scales $\tau_\mathrm{m}=r_\mathrm{p}/\dot r_\mathrm{p}$
for two- (open symbols) and three-dimensional (filled circles) isothermal 
simulations. As in Fig. \ref{3Dfig6} we assume a disk mass of 5 \mjup.
The dotted line and the dashed line indicate the analytical results from
\cite{2002ApJ...565.1257T} for Type I migration in three-dimensional and 
two-dimensional disks, respectively. The horizontal dash-dotted line indicates
migration on the viscous time scale of the disk (Type II migration).}
\label{3Dfig7}
\end{figure}

We will now turn to the question whether the differences in flow structure
we find affect accretion and migration rates of embedded planets.
In Fig. \ref{3Dfig6} we show the measured accretion rates as a function of 
planetary mass. Note that in this section we do not limit ourselves to low-mass
planets in order to compare with previous numerical results 
\citep{2003ApJ...586..540D}. The two-dimensional results, shown by the open 
symbols, were discussed extensively in \cite{2006A&A...450.1203P} and we show 
them here just for comparison. We have modeled four different planetary masses,
spanning a range from deep in the linear regime ($M_\mathrm{p} = 0.6$ \mearth) 
to well in the non-linear, gap-opening regime ($M_\mathrm{p} = 1.0$ \mjup).

Comparing the results of two- and three-dimensional simulations for the two
planets with the highest mass, we find very good agreement. Especially for the
1 \mjups planets this is to be expected, because the Roche lobe of this planet
is larger than the scale height of the disk, which makes the two-dimensional
approximation applicable. But also for the planet of $0.1$ \mjups we find
good agreement, which contradicts the result of \cite{2003ApJ...586..540D}.
However, this discrepancy is small: for this planet, \cite{2003ApJ...586..540D}
find a difference of approximately a factor of two between 2D and 3D results,
while we find a difference of $1.3$. It is possible that because our method
is specifically designed to handle shocks in a correct way, non-linear effects
start to play a role for slightly lower planetary masses. Our results on 
migration rates support this suggestion (see below).

Because the Roche lobes of the low-mass planets are much smaller than the disk 
thickness the two-dimensional approximation breaks down. This affects migration
as well as accretion, but the effect on the accretion rates is the most
dramatic. When in the two-dimensional approach mass is taken away from the 
location of the planet, what essentially happens is that a whole column 
of gas is accreted onto the planet. In the full three-dimensional problem this does not
happen, and therefore the accretion rates are significantly lower. For the
smallest planet we consider ($M_\mathrm{p} = 0.6$ \mearth) the difference 
amounts to more than three orders of magnitude (see the right panel of Fig. 
\ref{3Dfig6}). 

Qualitatively our results agree very well with the results of 
\cite{2003ApJ...586..540D}. Interestingly, \cite{2003MNRAS.341..213B} do not
find an increase in growth time towards the low-mass end, but this is 
probably due to a lack of resolution. Our results show that in order to capture
significant amounts of gas the mass of a solid core needs no be at least 
several Earth masses.

Migration rates also differ between 2D and 3D disks 
\citep{2002ApJ...565.1257T}. Again, the difference is most pronounced for 
low-mass planets, while for planet with masses compared to Jupiter the 
two-dimensional approximation gives valid results. In Fig. \ref{3Dfig7} we 
show the migration time scales as a function of planetary mass. For a planet on
a circular orbit, the migration time scale is related to the torque in the
following way:
\begin{equation}
\frac{1}{\tau_\mathrm{m}}=\frac{\left|\dot r_\mathrm{p}\right|}{r_\mathrm{p}}=
\frac{2 \left|\dot L\right|}{L}=\frac{2 \left|\mathcal{T}\right|}{L},
\end{equation}
where $L$ is the angular momentum of the planet and $\mathcal{T}$ is the 
torque, which is what we measure during the simulations.

\begin{figure*}
\includegraphics[width=17cm]{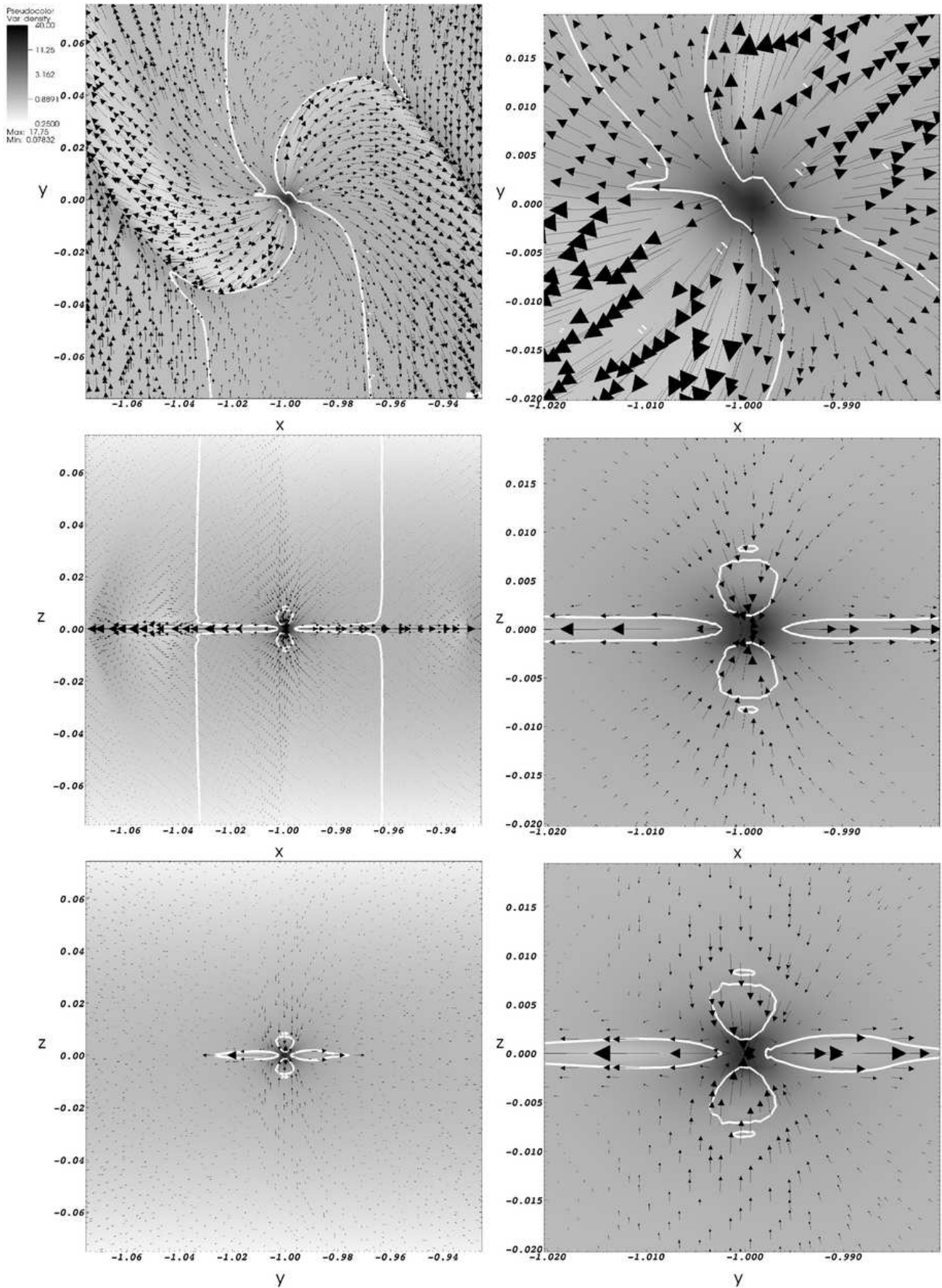}
\caption{Same as Fig. \ref{3Dfig2}, but for a nearly isothermal model with 
$\Gamma=1.01$.}
\label{3Dfig8}
\end{figure*}

\begin{figure*}
\includegraphics[width=17cm]{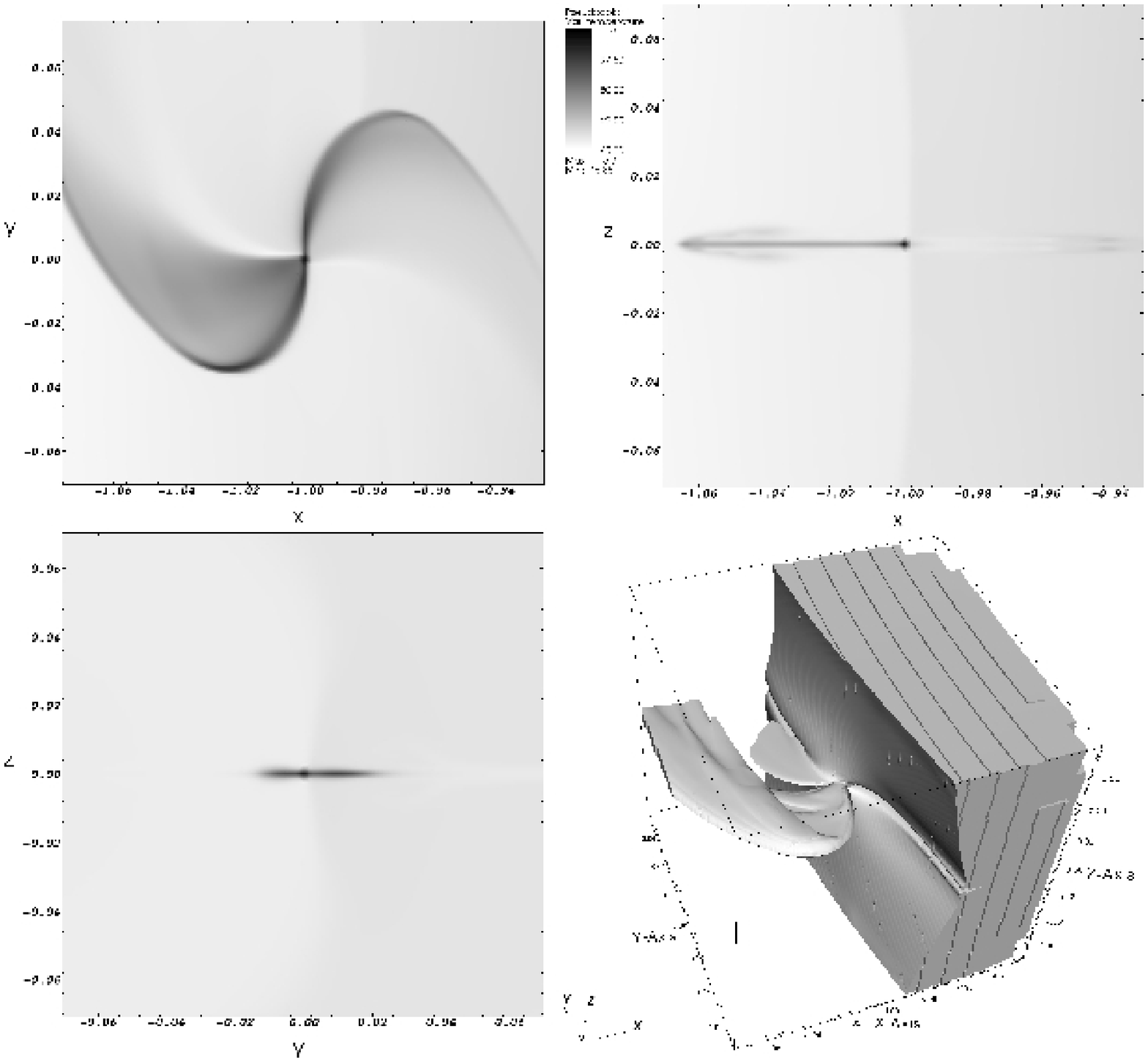}
\caption{Temperature structure around a 5 \mearths planet after 10 orbits 
for $\Gamma=1.01$. Top left panel: slice through $z=0$ (equatorial plane).
Top right panel: slice through $y=0$ ($\phi=\pi$). Bottom left panel: slice 
through $x=-1$ ($r=1$). Bottom right panel: isovolume indicating a temperature
higher than the initial temperature at the location of the planet.}
\label{3Dfig9}
\end{figure*}

Again, for the 1 \mjups planet the 2D result agrees very well with the 3D 
result, for the same reason as the accretion rates. In fact, they approach the
migration rate dictated by the viscous evolution of the disk (Type II 
migration). On the low-mass end, we can nicely reproduce the analytical results
from \cite{2002ApJ...565.1257T} for Type I migration in 2D as well as in 3D.
In this regime, 3D migration time scales are approximately a factor of two
larger than for 2D disks. 

In between the two regimes, we find that for a $0.1$ \mjups planet 2D and
3D results differ significantly, almost an order of magnitude. This is again
in contradiction with \cite{2003ApJ...586..540D}, but note that the main
difference lies in the 2D results. These differences were already discussed 
in \cite{2006A&A...450.1203P} and we will not repeat that discussion here. We only note that
the results for a three-dimensional disk agree very well with 
\cite{2003ApJ...586..540D}. Also, the departure from the Type I analytical 
result for the $0.1$ \mjups planet indicates that for this planet non-linear
effects start to play a role, in accordance with the suggestion made above 
that these effects are also visible in the accretion rates.

Summarizing, we find excellent agreement with analytical results of torques in 
the low-mass regime as well as in the high-mass regime. In between, the results
are not clear but \cite{2005MNRAS.358..316D} showed that this may be a general 
problem for this transition regime. The measured accretion rates agree very 
well qualitatively with \cite{2003ApJ...586..540D}, again with the exception
of the transition regime between linear and non-linear interaction with the 
disk.

\subsection{Nearly-isothermal models}
As a bridge between the simple, locally isothermal models of the previous 
section and the full radiation-hydrodynamical simulations we now consider 
models that do include the energy equation, but in which we treat the cooling
in a simplified way. Consider the relation between internal energy and pressure
for a perfect gas:
\begin{equation}
p=\left(\Gamma-1\right)\epsilon,
\end{equation}
which shows that the adiabatic exponent $\Gamma$ is a measure of how a change
in internal energy affects the gas pressure, and, given the density, therefore
also the temperature of the gas. For the isothermal case, $\Gamma=1$, 
pressure and internal energy evolve independently and the gas can be compressed
indefinitely without changing the temperature. To simulate cooling in a very
simplified way one can take a value of $\Gamma$ very close to 1, which was
done by \cite{2003ApJ...589..556N,2003ApJ...589..578N} for the two-dimensional 
planet-disk problem. In \cite{2006A&A...450.1203P} we commented already that 
this approach
affects planetary accretion rates to a large extent, because even for 
$\Gamma=1.001$ the planet forms a hot bubble around itself that prevents matter
from entering the Roche lobe.

In Fig. \ref{3Dfig8} we show the density and velocity structure around a
5 \mearths planet after 10 orbits. This figure can be directly compared to
the isothermal model depicted in Fig. \ref{3Dfig2}. The global properties
of the velocity field are similar to the isothermal case: inflow from above
and below the planet and outflow in the midplane of the disk. Both inflow
and outflow are again supersonic.

The central density is approximately 2 times lower in the nearly-isothermal
model compared to the fully isothermal case. This is due to the fact that
it is not only the density gradient that determines the pressure gradient,
but also the temperature. This means that part of the pressure-support for
the planetary envelope comes from the temperature gradient. We will see
below that this lower central density substantially affects the accretion rate onto the 
planet.

Another important difference that can be seen in the top left panels of Figs.
\ref{3Dfig2} and \ref{3Dfig8} is the change in outflow structure. While in the
isothermal case of Fig. \ref{3Dfig2} both outflow streams appear to be symmetric
this is no longer the case when changes in temperature are taken into account.
In the top right panel of Fig. \ref{3Dfig8} we see that the outflow towards
positive $y$ appears to be much more collimated along the $y$-axis than the 
outflow towards negative $y$, which is responsible for the asymmetry seen in 
the top left panel of Fig. \ref{3Dfig8}. It is important to realize at this point
that these asymmetries may suffer from resolution effects. Although in these 
simulations we resolve the Roche lobe by approximately 50 cells, the 
high-density region inside of the outflow is only resolved by 5 cells. 
Resolution studies show that the direction of the outflow does not depend on
resolution, but the magnitude of the outflow does. This is because the central
temperature and density also scale with resolution, as the minimum distance
to the planet does, and therefore also the depth of the potential well.
The equatorial outflow therefore critically depends on the conditions near
the center of the planet where the density and the temperature reach their 
highest values. Therefore there exists a connection between the conditions
deep in the atmosphere of the planet and the planet's immediate surroundings.

In Fig. \ref{3Dfig9} we show the temperature structure around the same planet as
in Fig. \ref{3Dfig8}. In this model, the central temperature has risen from 
50 K initially to more than 100 K. Note that it is only a very small region
that acquires this higher temperature. This reflects the fact that the 
accreting region is only 10 \% of the planet's Roche lobe. In this region,
the gas is heated by the release of potential and kinetic energy of the
accreted gas. The two streams of outflowing material
along the $y$-axis are clearly visible due to their high temperature. Also 
the collimation of the stream in the positive $y$-direction is apparent in
the top left panel of Fig. \ref{3Dfig9}. The outflows remain confined to the
midplane of the disk until they reach the spiral waves induced by the planet
that are responsible for upward motion of gas. However, for a low-mass planet 
this upward motion is not strong, and from the top right panel of Fig. 
\ref{3Dfig9} we see that the hot flow does not reach high altitudes. 

\begin{figure}
%\resizebox{\hsize}{!}{\includegraphics[bb=250 10 500 245]{paardekooperfig10.eps}}
\resizebox{\hsize}{!}{\includegraphics[]{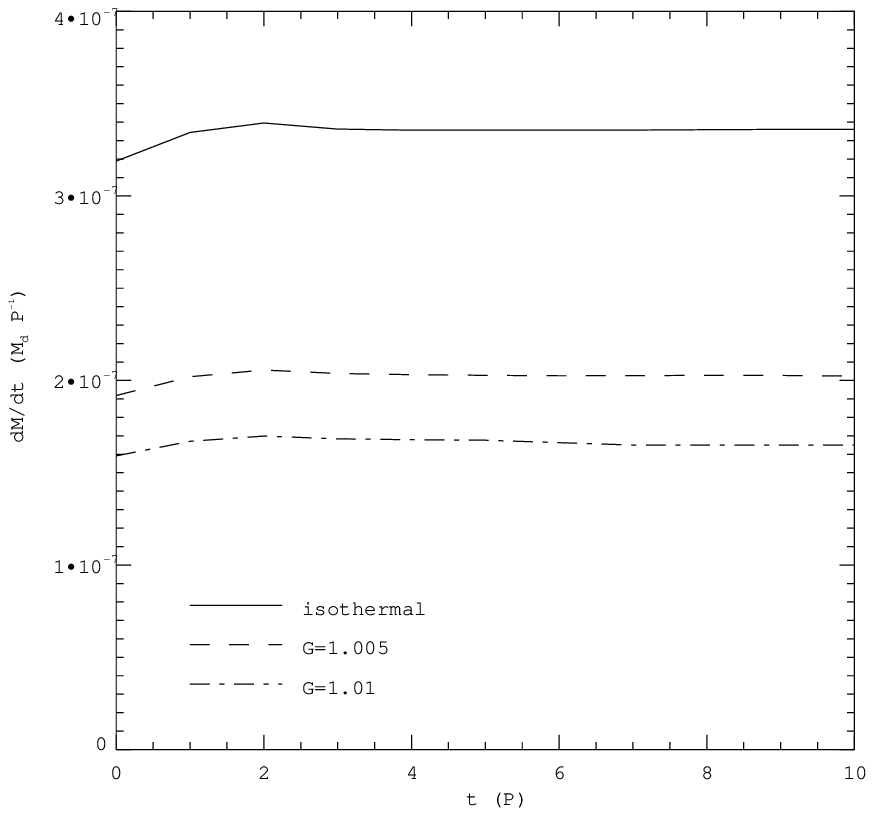}}
\caption{Accretion rates onto a 5 \mearths planet for three different 
equations of state.}
\label{3Dfig10}
\end{figure}

Another important effect can be seen in the bottom right panel of Fig. 
\ref{3Dfig9} that shows the three-dimensional temperature distribution around 
the planet. The solid volume indicates temperatures higher than the initial 
temperature at the location of the planet. The large solid structure at
$x>-1$ is due to the global temperature gradient. To the left of this structure
we see the effects of the planet, especially the spiral shaped hot outflow.
But note that the corotational region located at $x=-1$ is also asymmetrically
heated: for positive values of $y$ the corotation region is locally heated 
by the planet, while this does not happen for negative $y$. Again, this is 
a consequence of the asymmetric hot outflow that originates deep inside the
envelope of the planet. When the pressure remains the same, however, this
temperature distribution leads to an asymmetric density distribution which will
affect the total torque on the planet. Because the region \emph{behind}
the planet is heated compared with the region in front of the planet, the 
density will be relatively low behind the planet which leads to a torque
that will be \emph{positive} if the effect is strong enough. We will see below 
that for this model indeed the total torque on the planet is positive.

Before looking at the torques onto the planet we consider the measured 
accretion rates. In Fig. \ref{3Dfig10} we show the accretion rate as a function
of time for three different equations of state: the isothermal result
discussed in Sect. \ref{3DsecIso} together with two nearly-isothermal models. 
It turns out that the accretion rate depends sensitively on the central 
temperature. This is not surprising, because the pressure in near the center
of the planet is determined mainly by the potential of the planet, which 
remains the same. Therefore, when the central temperature rises by a factor
of 2, the central density will be a factor of 2 lower which in its turn
affects the accretion rate onto the planet. From Fig. \ref{3Dfig9} we saw that
for $\gamma=1.01$ the central temperature indeed was a factor 2 higher than
the initial temperature, while from Figs. \ref{3Dfig2} and \ref{3Dfig8} we see
that the central density is lower by approximately a factor 2 for the model 
with $\Gamma=1.01$ compared to the isothermal model. In Fig. \ref{3Dfig10} we
see that this change in density is reflected in an accretion rate that is 
lowered again by a factor 2. 

\begin{figure}
%\resizebox{\hsize}{!}{\includegraphics[bb=260 10 500 245]{paardekooperfig11.eps}}
\resizebox{\hsize}{!}{\includegraphics[]{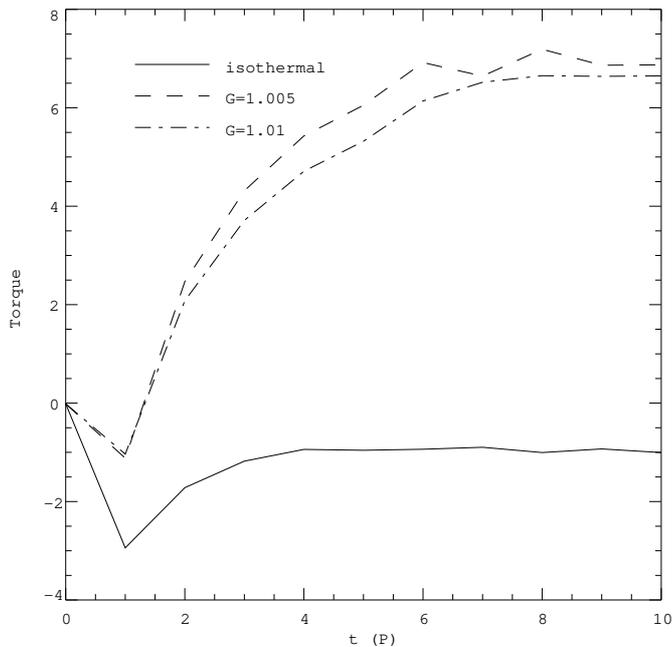}}
\caption{Total torque onto a 5 \mearths planet for three different equations 
of state. The torques are normalized to the absolute value of the analytical 
result of \cite{2002ApJ...565.1257T}.}
\label{3Dfig11}
\end{figure}

\begin{figure}
%\resizebox{\hsize}{!}{\includegraphics[bb=255 10 500 245]{paardekooperfig12.eps}}
\resizebox{\hsize}{!}{\includegraphics[]{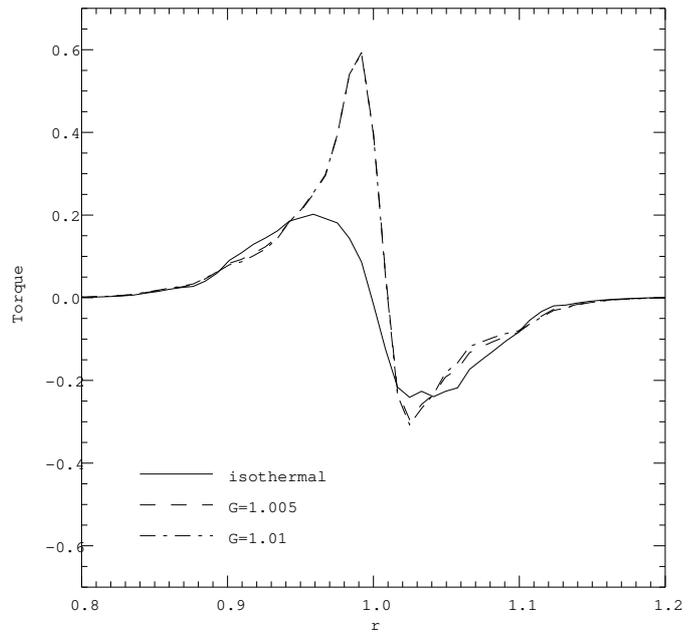}}
\caption{Radial torque profile for a 5 \mearths planet for three different 
equations of state. The torques are normalized in such a way that the total
torque for an isothermal equation of state is equal to $-1$.}
\label{3Dfig12}
\end{figure}

When we take $\Gamma$ closer and closer to 1, we expect to retrieve the
isothermal results. However, as was already mentioned in 
\cite{2006A&A...450.1203P}, even for $\Gamma$ extremely close to 1 the central
temperature of the planet rises significantly. In fact, due to the necessary
divisions by the factor $\Gamma-1$ \citep[see][]{1995A&AS..110..587E} one has 
to worry about numerical stability for $\Gamma \approx 1$. Therefore the lowest
value of $\Gamma$ we considered is $\Gamma=1.005$. For this model, the central
temperature rises to approximately 90 K, which is again reflected in a lower
accretion rate than for the isothermal model (see Fig. \ref{3Dfig10}).

In Fig. \ref{3Dfig11} we show the total torque onto the planet for the same
models as in Fig. \ref{3Dfig10}. The torques are normalized to the absolute
value of the analytic result of \cite{2002ApJ...565.1257T}, and as was already
shown in Fig. \ref{3Dfig7} for the isothermal model we find very good agreement
with the analytical result. This can not be said of the nearly-isothermal 
models, however. Indeed, for both values of $\Gamma$ we considered we find
a \emph{positive} torque, indicating outward migration. But not only is the
direction reversed, also the magnitude of the torque is surprising: more than
six times larger than for the isothermal case. The actual value of $\Gamma$
has no influence on this result. Note that it takes a few orbits longer for
the nearly-isothermal models to reach a steady state with respect to the 
torque, which was not the case for the accretion rates. This is probably due
to the fact that the torque is modified by the equatorial outflow. This
material has to come close to the planet which takes a few orbits.

It is interesting to find out where exactly this large positive contribution
to the torque originates. In order to answer this question, we show in Fig. 
\ref{3Dfig12} the radial torque profile for the three different equations of
state. For the isothermal case we see the usual 
\citep[see][]{2003MNRAS.341..213B,2003ApJ...586..540D} contributions from
the inner and the outer spiral wave, where the inner wave exerts a positive
torque on the planet and the outer wave a negative torque. It is clear that
even from inspection by eye that the torque due to the outer wave is slightly
stronger which leads to inward migration. 

The situation is totally different for nearly-isothermal models, as can be 
seen in Fig. \ref{3Dfig12}. For the outer disk the changes are marginal, which
is not surprising because the structure of the outer wave in Figs. \ref{3Dfig2}
and \ref{3Dfig8} appears very similar. This in contrast with the inner wave,
which has a much more open structure for the nearly isothermal model compared
with the truly isothermal model. This changing wave structure leads to a net
positive torque on the planet. 

From the temperature plots in Fig. \ref{3Dfig9} we conclude that the outflowing
material originates deep within the Roche lobe of the planet. This is an 
indication that accretion and migration rates of forming planets are closely 
linked. This was already suggested by \cite{2003ApJ...586..540D} for the 
isothermal case. However, only when the energy equation is taken into account
the migration behavior changes this drastically. Although in this section
we have treated cooling in a simplified way, it provides useful insight in 
what we may expect from the full radiation-hydrodynamical models. In the next 
section, we will consider a more realistic treatment of the energy balance by 
explicitly including radiative cooling. 

\begin{figure*}
\includegraphics[width=17cm]{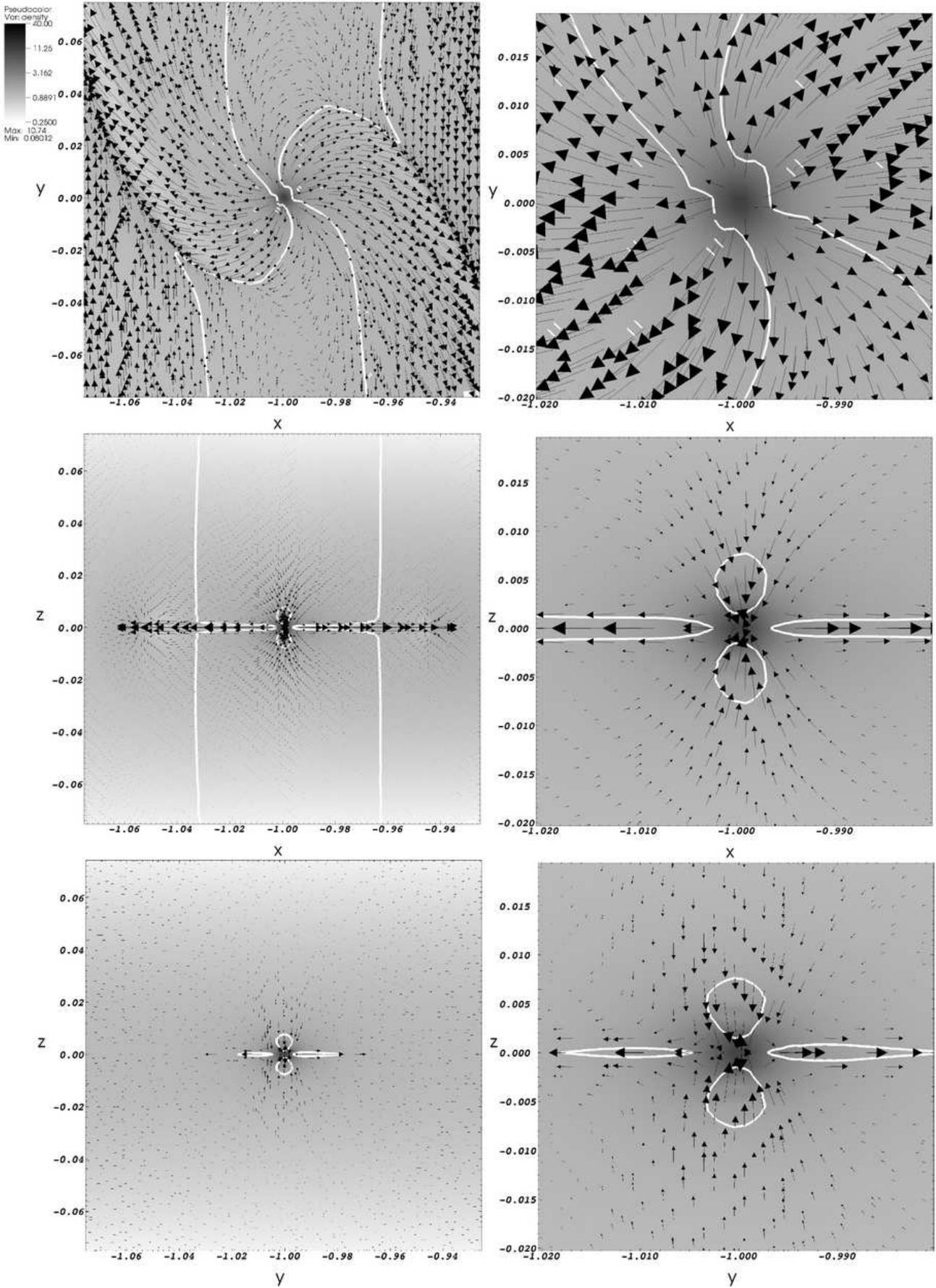}
\caption{Same as Fig. \ref{3Dfig2}, but for a non-accreting
radiation-hydrodynamical model with $\rho_0=10^{-11}~\mathrm{g~cm^{-3}}$.}
\label{3Dfig13}
\end{figure*}

\begin{figure*}
\includegraphics[width=17cm]{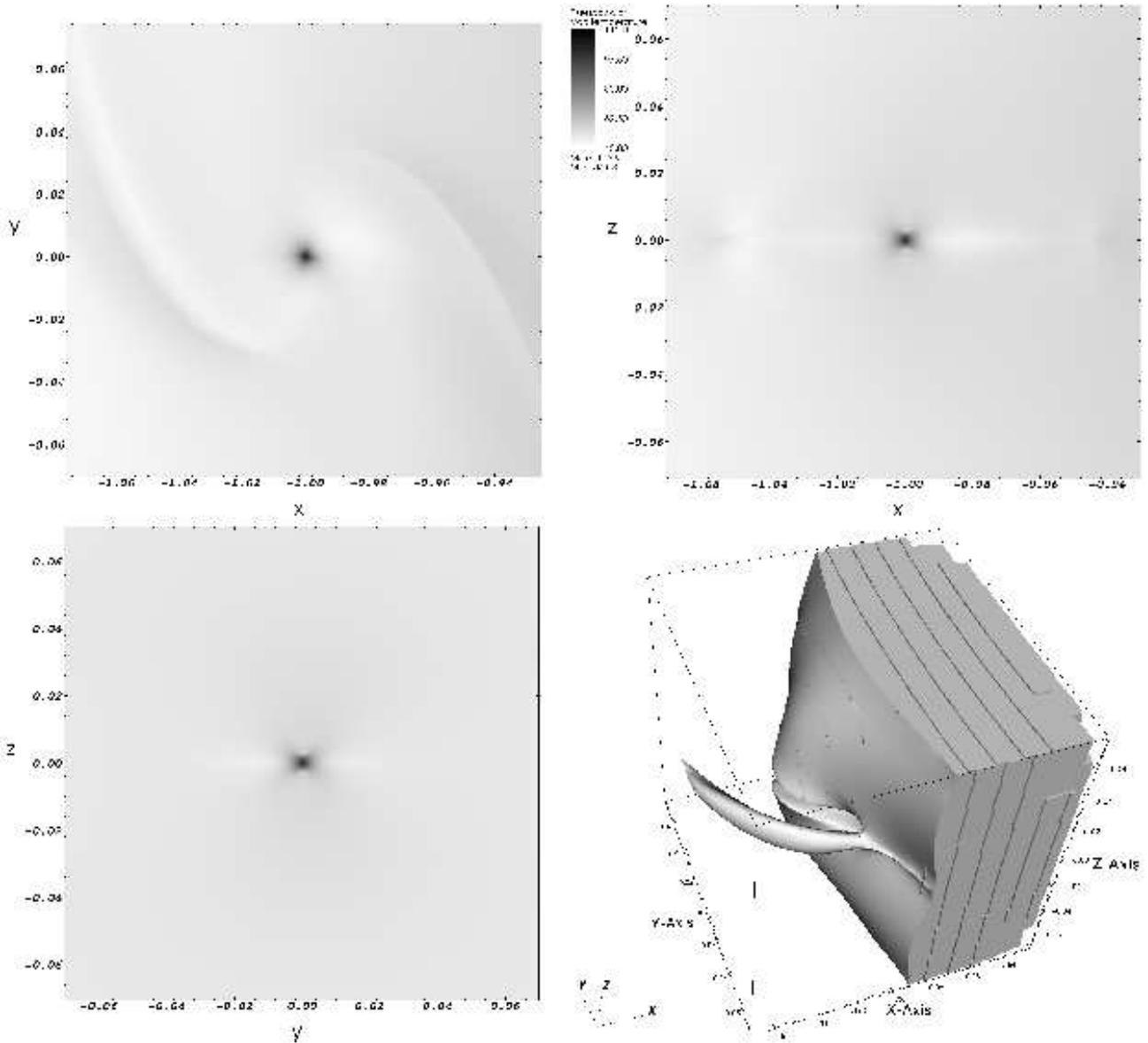}
\caption{Same as Fig. \ref{3Dfig9}, but for a non-accreting 
radiation-hydrodynamical model.}
\label{3Dfig14}
\end{figure*}

\begin{figure*}
\includegraphics[width=17cm]{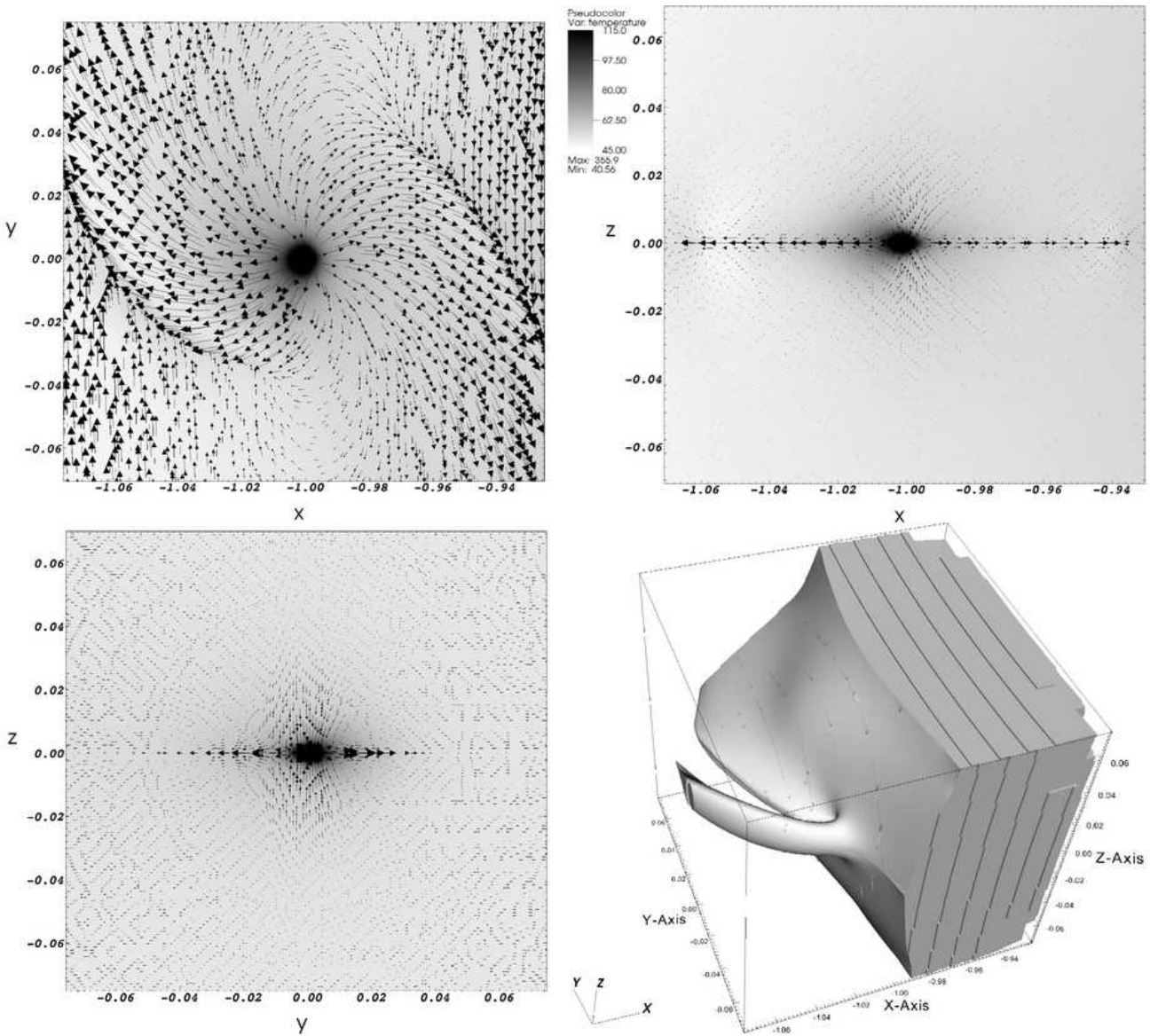}
\caption{Same as Fig. \ref{3Dfig9}, but for an accreting 
radiation-hydrodynamical model. Black arrows indicate the velocity field.}
\label{3Dfig15}
\end{figure*}

\subsection{Radiation-hydrodynamical models}
The full radiation-hydrodynamical models differ from the nearly-isothermal
models in two ways: first of all, the adiabatic exponent is set to 
$\Gamma=1.4$, which reduces the compressibility of the gas. Second, the FLD
module calculates the energy diffusion through radiation. Therefore, the 
cooling properties of the gas depend on the local temperature and density.
When cooling is not efficient, we expect that due to the higher value of
$\Gamma$ the effects seen in the nearly-isothermal models will be enhanced
even more. This will certainly happen in the high density envelope of the
planet, and because we saw in the previous section that temperature effects
deep within the Roche lobe of the planet affect the migration behavior we
may expect to see similar things in the radiation-hydrodynamical models.

In order to distinguish the effects from the accretion itself from the 
overall radiation-hydrodynamical behavior of the disk, we consider two cases
in this section: an accreting 5 \mearths planet and a non-accreting planet of 
the same mass. The non-accreting planet is allowed to build a high-density
atmosphere because no mass is removed from the Roche lobe. However, as was
already shown by \cite{2003ApJ...586..540D} for the isothermal case, even
in this case the envelope does not reach a state of hydrostatic equilibrium
in the vertical direction. Material still rains down into the Roche lobe
from above and below the planet, some of which subsequently flows out again.
Thus even in the case of a non-accreting planet there may exist a close
connection between the state of the material deep within the planetary envelope
and the torques due to material outside the Roche lobe.

In Fig. \ref{3Dfig13} we show the density and velocity structure for the 
non-accreting planet of 5 \mearth. The color scale is the same as in Figs. 
\ref{3Dfig2} and \ref{3Dfig8}, which tells us immediately that the density of
the planetary envelope is very low compared to the isothermal and 
nearly-isothermal cases. This reflects the effect of the lower compressibility
of the gas. We will see below that this has dramatic effects on the ability
of the planet to accrete gas from the disk.

Looking at the velocity field in Fig. \ref{3Dfig13} we see that indeed there is
no hydrostatic equilibrium in the vertical direction. The same flow pattern 
appears as in the isothermal and nearly-isothermal cases: inflow from above
and below and outflow along the equator of the planet. Inflow velocities are
lower in this case because the planet does not accrete matter 
\citep{2003ApJ...586..540D}. The outflow appears much more symmetric in all
directions compared to the previous nearly-isothermal case. Therefore 
radiative cooling restores the isothermal situation in these regions of 
lower density than the planetary core. However, we will see that still 
important differences arise in the torque on the planet.

In Fig. \ref{3Dfig14} we show the temperature structure around the planet, on 
the same scale as in Fig. \ref{3Dfig9}. The most important difference with
the midplane temperature structure of the nearly-isothermal models is the
absence of the accretion jets because in Fig. \ref{3Dfig14} we show a 
non-accreting model. The temperature at the center of the planet is also lower;
about a factor of 2 higher than the initial temperature. This agrees with 
Eq. \eqref{3DeqTconv} for the maximum central temperature that can be reached 
in absence of convection. We can also identify the spiral waves in the
temperature structure, because when the gas is compressed in the wave the
temperature rises. In the dense midplane of the disk the gas can not radiate
away this energy efficiently, and therefore the spiral waves have a higher 
temperature. 

Inside the equatorial outflow the gas expands, and therefore cools down. This
effect causes the temperature close to the hot core to be lower than the 
initial temperature. These regions show up as dark areas in Fig. \ref{3Dfig14},
most notably in the top right panel. 

Another difference with the nearly-isothermal models can be spotted most 
clearly in the bottom-left panel of Fig. \ref{3Dfig14}. While the heat generated
in the envelope of the planet remains confined to the midplane of the disk
in the nearly-isothermal models (see Fig. \ref{3Dfig9}), in the 
radiation-hydrodynamical models it can diffuse to higher altitudes. Because
of the large opacity gradient in the vertical direction, this is the 
preferable direction for radiative diffusion.

In the bottom right panel of Fig. \ref{3Dfig14} we again, just as in Fig. 
\ref{3Dfig9}, show a 3D isovolume of the temperature structure around the 
planet. It is clear that also the spiral wave structure in the temperature
extends further in the vertical direction due to radiative heat diffusion. 
Also it is clear that the strong accretion jets are absent. However, what 
remains is the asymmetric temperature structure in the corotation region.
There is a bump visible for positive $y$ ($\phi<\phi_\mathrm{p}$, the 
region behind the planet) while there is a small hole for negative $y$. 
This asymmetric corotational temperature structure affects the corotation
torque.

\begin{figure}
%\resizebox{\hsize}{!}{\includegraphics[bb=250 10 500 245]{paardekooperfig16.eps}}
\resizebox{\hsize}{!}{\includegraphics[]{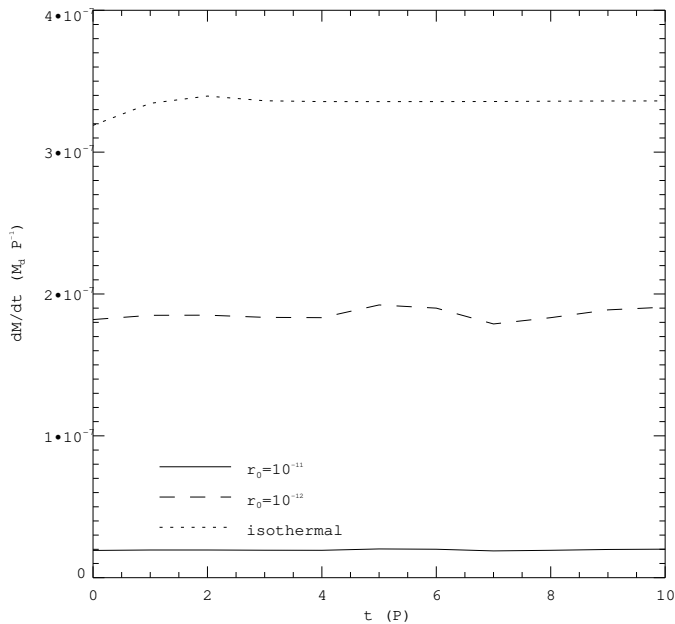}}
\caption{Accretion rates onto a 5 \mearths planet for radiation-hydrodynamical
models with two different densities. For comparison, the isothermal accretion
rate is also shown.}
\label{3Dfig16}
\end{figure}

In Fig. \ref{3Dfig15} we show the temperature and velocity structure for the case of an accreting planet. It is immediately clear that the temperature is raised considerably in the planetary envelope due to the accretion proces. 
Note that in this case the planetary envelope is unstable to convection 
according to Eq. \eqref{3DeqTconv}: the heat generated in the accretion process
can not be transported to the surface of the Hill sphere by radiation alone.
Although Eq. \eqref{3DeqTconv} is of limited applicability because of the 
assumption that the local pressure scale height is the same as for the total
disk, we can identify heat transport by fluid motion in the equatorial
plane of the disk. The top left panel of Fig. \ref{3Dfig15} shows that the
outflow velocity pattern is changed with respect to the non-accreting model:
it seems that heat flows preferentially into the outer disk. However, it
should be stressed that this observed asymmetry may be caused by insufficient
resolution at the very center of the planetary envelope. Nevertheless, 
these hot outflows cause significant changes in the temperature and density
distribution in the immediate surroundings of the planet and are therefore
potentially important for the migration behavior of the planet. The details
of the accretion process should therefore be subject of further study.

Also from Fig. \ref{3Dfig15} we see that the convective flow is confined to 
the midplane of the disk: vertical velocities in the top right and bottom
left panel are all directed towards the planet. This shows that the vertical 
thickness of the hot bubble surrounding the planet is due to radiative heat
diffusion. Its vertical extension is roughly the size of the Hill sphere of
the planet.

The formation of the hot bubble has dramatic consequences for the measured
accretion rate onto the planet. In Fig. \ref{3Dfig16} we show the accretion
rate as a function of time for the isothermal case (dotted line) and two
radiation-hydrodynamical models with two different initial values for the 
density at the location of the planet. The solid line indicates an initial 
density of $\rho_0=10^{-11}~\mathrm{g~cm^{-3}}$ which is our nominal value.
For this density we measure an accretion rate that is more than an order of 
magnitude lower than for the locally isothermal equation of state. The hot
bubble surrounding the planet severely limits the gas flow towards the planet.
Because this bubble is created by the accretion process itself, accretion is
a self-limited process in this case. 

\begin{figure}
%\resizebox{\hsize}{!}{\includegraphics[bb=260 10 500 245]{paardekooperfig17.eps}}
\resizebox{\hsize}{!}{\includegraphics[]{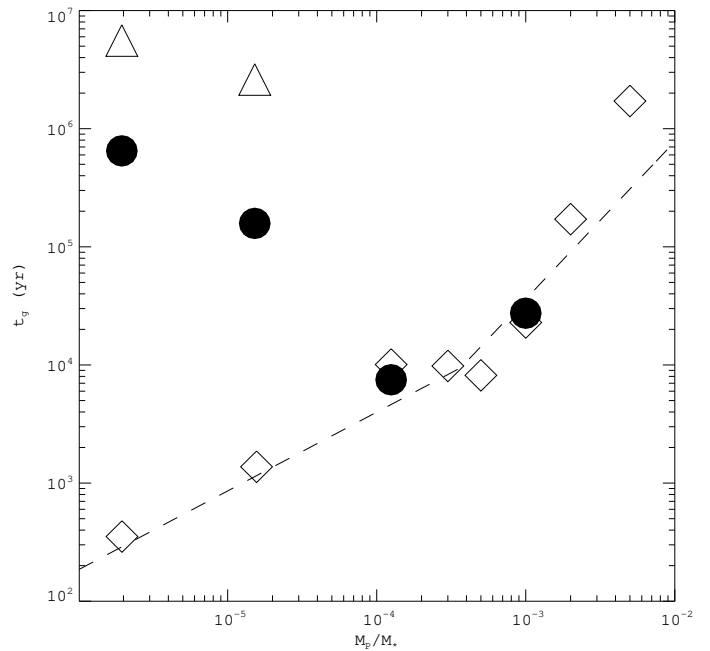}}
\caption{Same as the right panel of Fig. \ref{3Dfig6}, with the 
radiation-hydrodynamical results (with our nominal initial density of $\rho_0=
10^{-11}~\mathrm{g~cm^{-3}}$) included as open triangles.}
\label{3Dfig17}
\end{figure}

For a lower density the cooling efficiency increases. In fact, according to Eq.
\eqref{3Deqtcool} the cooling time is proportional to the square of the 
density. 
For a model with a ten times lower density (dashed line in Fig. \ref{3Dfig16})
we measure an accretion rate that is almost ten times higher than for our 
nominal model. This accretion rate is comparable to the results for 
nearly-isothermal models (see Fig. \ref{3Dfig10}). Lowering the density even
more did not change the accretion rate significantly. This also agrees with
the nearly-isothermal models, where we saw that even for $\Gamma$ as low as
$1.005$ the accretion rate remained far from the isothermal result. This can
be understood in terms of the high central density required to achieve the
isothermal accretion rate. From Fig. \ref{3Dfig2} we see that the central density
is more than a factor 37 higher than the initial density. It is very hard to 
cool such a high-density envelope efficiently, especially when 
accretion-generated heat is considered as well. 

In Fig. \ref{3Dfig17} we show the growth time scales for the isothermal results
(see also Fig. \ref{3Dfig6}) with the radiation-hydrodynamical results added as
open triangles. The changes with respect to the isothermal models are not as 
dramatic as the changes from moving from 2D simulations to 3D simulations, but
still it represents another order of magnitude slower accretion. Also from
Fig. \ref{3Dfig17} we see that the effect is slightly less strong for a planet of
$0.6$ \mearth. This is because the potential for this planet is less deep,
but the decrease in accretion rate is still significant even for this very 
small planet.

\begin{figure}
%\resizebox{\hsize}{!}{\includegraphics[bb=260 10 500 245]{paardekooperfig18.eps}}
\resizebox{\hsize}{!}{\includegraphics[]{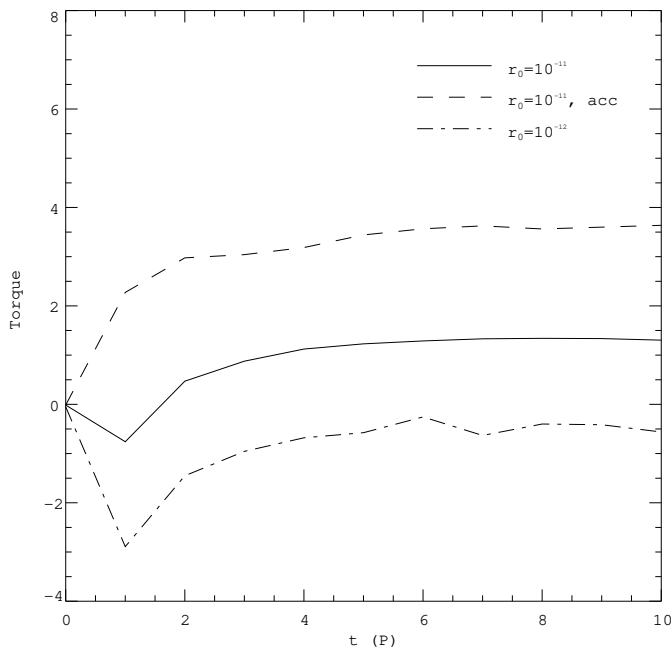}}
\caption{Total torque onto a 5 \mearths planet for three different radiation
hydrodynamical models. The torques are normalized to the absolute value of the 
analytical result of \cite{2002ApJ...565.1257T}.}
\label{3Dfig18}
\end{figure}

We now turn to the question of planetary migration within the 
radiation hydrodynamical models. From the nearly-isothermal models we know
that even the migration direction is very sensitive to temperature variations,
and we expect that because for our nominal density cooling is not efficient
we should recover at least some of the features of the nearly-isothermal
models.

In Fig. \ref{3Dfig18} we show the total torque acting on the planet as a function
of time, for three different radiation-hydrodynamical models. All torques reach
a constant value within 10 dynamical time scales, even though the cooling time
for the high-density case is somewhat longer. We checked that the torques did
not change over several cooling times. Again, as was the case for the 
nearly-isothermal models, the torque needs some more time to reach a steady 
value than in the isothermal case.

For our nominal, non-accreting model we observe that the total torque is again
positive, indicating outward migration. Its magnitude is comparable to the
ordinary Type I torque as calculated by \cite{2002ApJ...565.1257T}, and this
is much lower than the magnitude of the torque in the accreting, nearly 
isothermal models. It turns out that this is due to the accretion flow. Indeed,
for our accreting radiation-hydrodynamical model (dashed line in Fig. 
\ref{3Dfig18}) the magnitude of the torque increases by a factor of 3 compared
to the non-accreting case. It is the asymmetric, hot outflow that is 
responsible for a large part of the positive torque in the accreting model. 
Radiative heat diffusion tends to lower the magnitude of the torque, because
the flow is spread out more evenly around the planet. This effect is not
strong, however.

When we lower the density the magnitude of the torque decreases until at some
point it becomes negative. This is because the cooling time decreases with 
decreasing density, and therefore the disk is more and more able to cool
towards its initial temperature. From Fig. \ref{3Dfig18} we see that for a 
10 times lower density the total torque is negative, with a magnitude that
is approximately a factor 2 lower than the analytical Type I torque from
\cite{2002ApJ...565.1257T}. As we reduce the density even more the torque
approaches the analytical result even more, but the scatter around the 
equilibrium value increases. This is due to the fact that for such densities
the disk becomes very optically thin, which makes the problem less well suited
to treat in the diffusion approximation. 

For accreting models, we do not reach the Type I result, however. The results
on accretion indicated already that even for very low densities there is still
a very hot bubble formed around the planet, and the subsequent hot outflow
comes to dominate the torque, which we found to remain positive even for the
case of a 10 times lower density. This is again an indication that accretion 
and migration of embedded planets are closely linked. In order to make 
progress, the detailed process of accretion requires further study.

\begin{figure}
%\resizebox{\hsize}{!}{\includegraphics[bb=255 10 500 245]{paardekooperfig19.eps}}
\resizebox{\hsize}{!}{\includegraphics[]{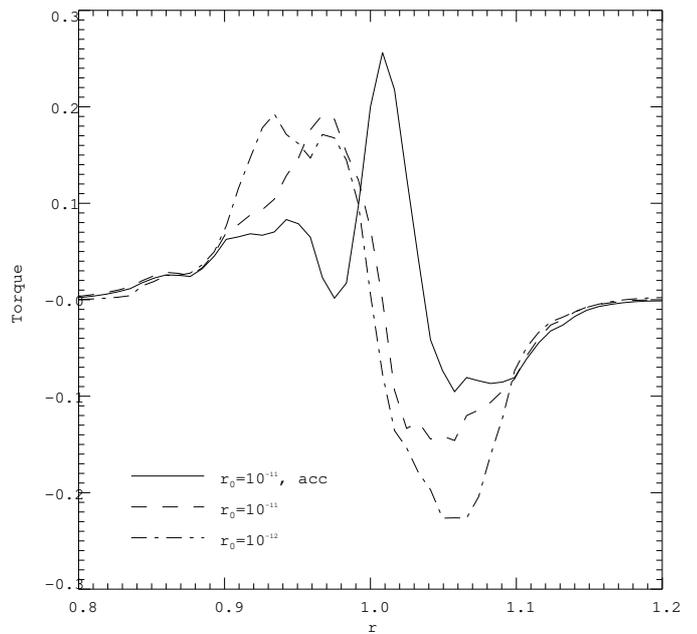}}
\caption{Radial torque profile for a 5 \mearths planet for three different 
radiation-hydrodynamical models. The torques are normalized in such a way that 
the total torque for an isothermal equation of state is equal to $-1$.}
\label{3Dfig19}
\end{figure}

Nevertheless, even for our non-accreting model we find a torque that is 
positive, which means that the hot accretion flow is not the whole story. 
It is interesting to see where in the disk this positive torque arises.
In Fig. \ref{3Dfig19} we show the torque on the planet as a function of radius. 
The low-density case (dash-dotted line), giving rise to a negative torque in 
Fig. \ref{3Dfig18}, agrees very well with the isothermal model (see Fig. 
\ref{3Dfig12}). For our nominal density (indicated by the dashed line in Fig.
\ref{3Dfig19}) we see that the magnitudes of the torques due to the inner and
the outer spiral wave are lower than for the low-density case. This is due
to compressional heating in the waves, as can be seen in the top left panel 
of Fig. \ref{3Dfig14}. Because of the higher temperature, the same pressure
distribution as in the isothermal case requires the density to be lower in the
waves, and therefore the torque is less strong. 

However, it turns out that the total torque due to the spiral waves is still
negative. This means that it is the corotation region that determines the sign
of the torque and therefore the direction of migration. In the next section, 
we will further specify where and how this torque arises using local, adiabatic
simulations.

Before we do that, we take a brief look at the accreting 
radiation-hydrodynamical model in Fig. \ref{3Dfig19}. The whole numerical set-up 
was the same as for our nominal density model in Fig. \ref{3Dfig19}, only now
accretion was turned on. All differences between the solid line and the dashed 
line in Fig. \ref{3Dfig19} arise from the accretion process. Note that the
sphere of influence of the accretion flow has a radius of approximately
$0.1~r_\mathrm{p}$, or $2~H$. The outflow of energy released by accretion 
heats the spiral waves even more, and consequently the torques decrease in
magnitude. On top of that, the asymmetric hot bubble seen in Fig. \ref{3Dfig15}
itself exerts a large positive torque onto the planet, which can be identified
as the large peak in the torque around $r=1$ in Fig. \ref{3Dfig19}. This peak
is less strong than in the nearly-isothermal case (see Fig. \ref{3Dfig12})
because the heat generated by accretion is diffused by radiation over a 
much larger volume.

%\begin{figure}
%\resizebox{\hsize}{!}{\includegraphics[]{fig20.eps}}
%\caption{Dependence of $\mathcal{T}_\mathrm{temp}$ on the radial
%  entropy gradient. Torques are normalised to the value obtained for
% $\mathcal{S}=-0.4$.}
%\label{3Dfig20}
%\end{figure}

\subsection{Local, adiabatic models}
In this section we take a step back and look at local, adiabatic simulations. 
That is, we keep $\Gamma=1.4$ but we switch off the radiative transport, and
we further reduce the computational effort by simulating only a small part
of the disk of 5 initial pressure scale heights around the planet. This volume
is large enough to capture the basic features of the flow, because the effects
of the low-mass planets we consider on the disk remain fairly local. We made
sure that the global resolution in these local models matches the highest local
resolution of the previously described models.

In these simplified simulations the gas is still heated by compression and
gravitational effects, while radiative cooling is absent. This way, these 
simulations would apply to a very dense region where the cooling time scale
is much larger than the dynamical time scale everywhere, also in the upper
layers of the disk. Although unrealistic for real protoplanetary disks, these
simulations allow us to pin down where the positive torque in a non-accreting
model arises while not being hindered by the cooling effects of radiation. 
These cooling effects will partially wash out the effects of compression, and
while the total torque is still positive, its origin is much harder to find.

In \cite{radlett}, it was found that in these models, a warm, underdense plume forms behind the planet, which is responsible for the torque reversal. We provide some additional discussion here. Further simulations have indicated that the torque due to temperature effects,
\begin{equation}
\mathcal{T}_\mathrm{temp}=\int_\mathrm{disk} \frac{\mathrm{G}M_\mathrm{p}\partial \rho}{|{\vec
    r} - {\vec r_\mathrm{p}}|^3} {\vec{r_\mathrm{p}}} \times ({\vec
    r}-{\vec r_\mathrm{p}}) {\vec d^3r},
\end{equation}
is proportional to the radial entropy gradient of the unperturbed disk:
\begin{equation}
\mathcal{T}_\mathrm{temp}\propto -\mathcal{S} \equiv -\frac{r}{S}\frac{dS}{dr}= -\frac{1}{T}\frac{dT}{dr}+\left(\Gamma-1\right)\frac{1}{\rho}\frac{d\rho}{dr}.
\end{equation}
In our model disk, $\mathcal{S}$ is negative, which gives a positive torque on the planet. Reversing the entropy gradient, one finds that the torque also changes sign. This is also seen using a different method \citep{clement}. We leave a detailed discussion of this effect, in particular its emergence from linear theory, to a future publication \citep{subm}. Here, we just note that the mechanism depends on the conservation of entropy along streamlines in the horseshoe region, which, in the presence of a background radial entropy gradient, leads to a strong corotation torque. In a barotropic disk, a similar mechanism operates involving the specific vorticity \citep[see][]{1991LPI....22.1463W}. It is interesting to note that, in contrast to the accretion rates, the torque displays discontinuous behavior at $\Gamma=1$. For an adiabatic disk, where density and temperature decrease outward, according to Eq. \eqref{eqent} the entropy gradient reaches a minimum for $\Gamma \rightarrow 1$, leading to a strong positive corotation torque. For $\Gamma=1$, however, we have a locally isothermal disk in which entropy is not conserved along streamlines, which kills the entropy-related corotation torque.

\section{Discussion}
\label{3DsecDisc}
Our results on gas accretion onto low-mass protoplanets suggest that mass
flow into the Hill sphere depends critically on the temperature profile
of the planetary envelope. In isothermal simulations, pressure gradients 
correspond to density gradients alone, and therefore a pressure-supported
isothermal envelope will have a very high central density. This leads to 
the relatively high accretion rates in isothermal models. When the energy 
balance is taken into account, part of the pressure gradient that is needed
to sustain the envelope is carried by a gradient in temperature and therefore
the corresponding density profile will be less peaked, and therefore the
measured accretion rate will go down. Moreover, the energy released in the 
accretion process raises the central temperature even further, slowing down
the mass flow even more. This way, accretion is a self-limiting process, 
and the subsequent accretion rate is more than an order of magnitude below
the isothermal result. The values we find are comparable to the solid accretion rates found by \cite{1996Icar..124...62P}, depending on the mass of the planet. Heating due to this planetesimal bombardment should lead to a similar drop in the accretion rate as observed here. We conclude that accretion rates obtained from isothermal simulations strongly overestimate the growth rate of embedded planets.

Because the release of accretion energy plays such a significant role in
accretion, migration and observations of protoplanetary envelopes
\citep{2006A&A...445..747K} it is important to consider the validity of
this accretion recipe. Although at first sight it seems a good idea to 
conserve total energy and momentum during accretion, it is not entirely
clear if this is indeed the case. The momentum carried by the accreted matter,
for example, might be transferred to the planet itself, which may induce
rotation of the protoplanetary core. 

As an example for an alternative to energy-conserving accretion, consider an 
accretion recipe that conserves entropy. Because at constant entropy the 
pressure is related to the density as:
\begin{equation}
p=K~\rho^\Gamma,
\end{equation}
we can, using the ideal gas law, immediately write down that
\begin{equation}
T=C~\rho^{\Gamma-1},
\end{equation}
for some constant $C$. This means that taking away mass from the grid actually
\emph{cools} the accretion region. In this case the assumption is that the
accretion energy is transferred to the planetary core. Note that in using
an isothermal equation of state, taking away mass from the grid amounts to
conservation of entropy as well. Because accretion plays such a major role
in non-isothermal planet-disk interaction, obviously in the growth of the 
planet but also in the migration behavior, more work on the details of the
accretion process is necessary.

But even for non-accreting planet, a hydrostatic envelope is never reached
\citep[see also][]{2003ApJ...586..540D}. Mass constantly flows in and out
of the Roche lobe, and it is not yet clear what role these outflows play
in planet-disk interaction. In our simulations we have found that the 
outflow is predominantly equatorial, irrespective of whether the planet
is accreting mass or not. The depth of the potential well does influence
the outflow, however, because it is driven by the high pressure region at
the location of the planet. We found that this outflow did not alter the 
torque balance in the isothermal limit, but because the gas expands rapidly
inside the outflow it affects the temperature in the midplane and it may
therefore be of importance in non-isothermal disks. 

The most interesting result from this study is the observed reversal of 
migration direction with respect to isothermal disks. We have tried to 
isolate the origin of the positive torque arising in adiabatic simulations,
and we have found that gas heating in the region where the horseshoe orbits
make their U-turn is the dominant effect. This heating is driven by a radial entropy gradient, where a negative gradient results in a positive torque on the planet.

Classical corotation torques tend to saturate when the viscous diffusion time scale 
across the librating region is longer than the libration time scale
\citep{2003ApJ...587..398O}. In our 
inviscid models, saturation should occur, but a much larger time range is 
needed to see the effect. The computational costs of radiation-hydrodynamical
simulations are too large to achieve this, however. Global adiabatic models
may provide more insight in this matter. The new torques arising from a radial entropy gradient may be kept unsaturated by energy transport by radiation, but only if the disk in radiative equilibrium has indeed a radial entropy gradient. More detailed disk models are needed to answer this question.

In our models where we solve the energy equation we have focused on a
5 \mearths planet. One run involved a $0.6$ \mearths accreting planet that
also moved outward, but more study is necessary to establish the trend, in particular whether it is a purely linear effect.

When the planet gets massive enough to substantially change the density in 
its direct environment, the situation changes. Due to the lower density near
the orbit of the planet the positive corotation torques decrease in strength
because the cooling efficiency increases. Note that the cooling time is
proportional to the density squared as long as the disk remains optically
thick. Therefore the outward migration induced by corotational heating will
quickly lose strength when the planet opens up a gap. For gap-opening planets,
this mechanism does not operate because the disk near the planet can cool
efficiently (density reductions of two orders of magnitude were measured in
2D simulations).

This new migration behavior of low-mass planets raises a few interesting
issues. First of all, because the direction of migration depends on the 
density, there exists a certain radius $R_0$ with density $\rho(R_0)$ where 
the total torque is zero. This radius is the same for all low-mass planets, because it depends on cooling properties of the disk only. Inside $R_0$, cooling is not efficient and planets
move outward. Outside $R_0$, planets move inward through ordinary Type I 
migration. Therefore all low-mass planets near $R_0$ in the disk will slowly 
approach $R_0$. In a simple power-law disk there is only one $R_0$, which,
in the Minimum Mass Solar Nebula is located around 10-15 AU. When the density
structure becomes more complicated (for example due to the presence of a 
gap-opening planet) there may be several locations in the disk where the total 
torque is zero. 

As the low-mass planets approach $R_0$, planet-planet interactions will 
eventually come to dominate. When this happens depends on the magnitude of 
the torque in that region. Interesting effects were observed for differentially
migrating low-mass planets by \cite{2006A&A...450..833C}. The planets may
lock into mean-motion resonances or be scattered out of the system.

As the disk evolves it slowly loses mass. Therefore, in a simple power-law 
disk, $R_0$ will shift inwards slowly. Once a planet has reached $R_0$ either
by outward or inward migration, it will slowly migrate inward due to the
dispersal of the disk. Therefore the migration time scale for low-mass
planets becomes directly coupled to the disk life time. While in the isothermal
Type I migration scenario migration may proceed on any time scale depending on
the disk mass, when the energy budget is taken into account this is no
longer the case. Low-mass planet migration will proceed at time scales
comparable to the disk life time, which is equal to the viscous time scale
of Type II migration. Therefore, this scenario of corotational heating 
provides a solution to the problem of fast-migrating cores. 

In using Eq. \eqref{3DeqOpacLow} in calculating the cooling time we have assumed
that the planet resides beyond the snow line. Interesting things may happen 
in regions where the opacity suddenly changes, as was pointed out by 
\cite{2004ApJ...606..520M}. Also, grain growth affects the opacity to a 
great extent. A disk in which all particles have grown to cm-size will
be optically thin, and therefore the mechanism of corotational heating will
not apply. However, such efficient growth of particles is not consistent
with observations of protoplanetary disks \citep{2005A&A...434..971D} and
therefore some sort of replenishment of small grains is needed. 

Another important effect that will have to be included is irradiation from
the central star. Although the disk is not severely perturbed by the 
low-mass planets considered in this paper, it is desirable to treat the
heating of the disk in a self-consistent way. However, detailed models
including irradiation from the central source are computationally very 
expensive and therefore hard to implement in a dynamical model. Flux-limited diffusion is not capable of handling possible shadowing effects properly, and therefore one has to switch to a ray-tracing method.  

This new migration behavior should in the end be included in population synthesis models of planet formation \citep{2004ApJ...604..388I,2004ApJ...616..567I,2005ApJ...626.1045I}. At the moment, these models require Type I migration to be reduced by a factor of a few in order to match the observed extrasolar planet population. The new insights in low-mass planet migration presented in this paper can increase the level of consistency in these models. 
  
\section{Summary and conclusions}
\label{3DsecCon}
In this paper we have presented the first global radiation-hydrodynamical 
simulations of the interaction between low-mass planets and protoplanetary 
disks. Specifically, we looked at accretion and migration rates in the mass 
regime of linear disk response. 

We have found that accretion and migration rates depend sensitively on the 
ability of the disk to radiate away energy generated by compression in 
the various flow regions around the planet. Compression deep within the 
planetary envelope leads to a decrease in accretion rate onto the planet
of more than an order of magnitude. Compression within the tidal waves
generated by the planet leads to a decreased Lindblad torque, which makes
the planet move inward but at a slower rate than predicted by analytical
models for isothermal Type I migration. Compression in the horseshoe region,
finally, leads to a large positive corotation torque that is able to
change the direction of migration, depending on the radial entropy gradient. 

All these effects depend critically on the cooling properties of the disk.
For a density appropriate for the Minimum Mass Solar Nebula at 5 AU we find
outward migration, while a ten times lower density gives rise to inward 
migration. Lowering the density even more we are able to reproduce the 
analytical Type I torque. This means that inward Type I migration is 
restricted to the outer parts of protoplanetary disks, or to older disks
that have a lower density. 

Because low-mass planet migration is driven by the decrease of density in 
the disk, it proceeds on the viscous time scale, just as Type II migration 
for gap-opening planets. This means that corotational heating is able 
to solve the problem of protoplanetary cores that move inward too fast to
form giant planets. All planets within the region where the cooling time
scale is much larger than the dynamical time scale will experience outward
migration, depending on the local entropy gradient.

Future studies should be aimed at finding the most appropriate recipe to 
simulate gas accretion onto the planet, and at including irradiation from
the central star to calculate the equilibrium disk temperature profile self-consistently. 

\begin{acknowledgements}
SP thanks Yuri Levin, Mordecai-Marc MacLow, Peter Woitke and John Papaloizou for useful 
discussions.
We thank Willem Vermin for his assistance at the Dutch National Supercomputer, and an anonymous referee for insightful comments that improved the quality of the paper. 
This work was sponsored by the National Computing Foundation
(NCF) for the use of supercomputer facilities, with financial support
from the Netherlands Organization for Scientific Research (NWO).
\end{acknowledgements}

\bibliographystyle{aa} 
\bibliography{8592.bib}

\begin{thebibliography}{59}
\expandafter\ifx\csname natexlab\endcsname\relax\def\natexlab#1{#1}\fi

\bibitem[{{Artymowicz}(1993)}]{1993ApJ...419..166A}
{Artymowicz}, P. 1993, \apj, 419, 166

\bibitem[{{Artymowicz}(2004)}]{2004ASPC..324...39A}
{Artymowicz}, P. 2004, in ASP Conf. Ser. 324: Debris Disks and the Formation of
  Planets, 39

\bibitem[{{Balbus} \& {Hawley}(1990)}]{1990BAAS...22.1209B}
{Balbus}, S.~A. \& {Hawley}, J.~F. 1990, \baas, 22, 1209

\bibitem[{{Barrett} {et~al.}(1994){Barrett}, {Berry}, {Chan}, {Demmel},
  {Donato}, {Dongarra}, {Eijkhout}, {Pozo}, {Romine}, \& {van der
  Vorst}}]{barrett}
{Barrett}, R., {Berry}, M., {Chan}, T.~F., {et~al.} 1994, {Templates for the
  Solution of Linear Systems: Building Blocks for Iterative Methods}
  ({Philadelphia, PA}: {SIAM})

\bibitem[{{Baruteau} \& {Masset}(2007)}]{clement}
{Baruteau}, C. \& {Masset}, F.~S. 2007, \apj, in press

\bibitem[{{Bate} {et~al.}(2003){Bate}, {Lubow}, {Ogilvie}, \&
  {Miller}}]{2003MNRAS.341..213B}
{Bate}, M.~R., {Lubow}, S.~H., {Ogilvie}, G.~I., \& {Miller}, K.~A. 2003,
  \mnras, 341, 213

\bibitem[{{Beckwith} \& {Sargent}(1996)}]{1996Natur.383..139B}
{Beckwith}, S.~V.~W. \& {Sargent}, A.~I. 1996, \nat, 383, 139

\bibitem[{{Bell} \& {Lin}(1994)}]{1994ApJ...427..987B}
{Bell}, K.~R. \& {Lin}, D.~N.~C. 1994, \apj, 427, 987

\bibitem[{{Cresswell} \& {Nelson}(2006)}]{2006A&A...450..833C}
{Cresswell}, P. \& {Nelson}, R.~P. 2006, \aap, 450, 833

\bibitem[{{D'Angelo} {et~al.}(2005){D'Angelo}, {Bate}, \&
  {Lubow}}]{2005MNRAS.358..316D}
{D'Angelo}, G., {Bate}, M.~R., \& {Lubow}, S.~H. 2005, \mnras, 358, 316

\bibitem[{{D'Angelo} {et~al.}(2002){D'Angelo}, {Henning}, \&
  {Kley}}]{2002A&A...385..647D}
{D'Angelo}, G., {Henning}, T., \& {Kley}, W. 2002, \aap, 385, 647

\bibitem[{{D'Angelo} {et~al.}(2003{\natexlab{a}}){D'Angelo}, {Henning}, \&
  {Kley}}]{2003ApJ...599..548D}
{D'Angelo}, G., {Henning}, T., \& {Kley}, W. 2003{\natexlab{a}}, \apj, 599, 548

\bibitem[{{D'Angelo} {et~al.}(2003{\natexlab{b}}){D'Angelo}, {Kley}, \&
  {Henning}}]{2003ApJ...586..540D}
{D'Angelo}, G., {Kley}, W., \& {Henning}, T. 2003{\natexlab{b}}, \apj, 586, 540

\bibitem[{{de Val-Borro} {et~al.}(2006){de Val-Borro}, {Edgar}, {Artymowicz},
  {Ciecielag}, {Cresswell}, {D'Angelo}, {Delgado-Donate}, {Dirksen}, {Fromang},
  {Gawryszczak}, {Klahr}, {Kley}, {Lyra}, {Masset}, {Mellema}, {Nelson},
  {Paardekooper}, {Peplinski}, {Pierens}, {Plewa}, {Rice}, {Sch{\"a}fer}, \&
  {Speith}}]{comparison}
{de Val-Borro}, M., {Edgar}, R.~G., {Artymowicz}, P., {et~al.} 2006, \mnras,
  370, 529

\bibitem[{{Dullemond}(2000)}]{2000A&A...361L..17D}
{Dullemond}, C.~P. 2000, \aap, 361, L17

\bibitem[{{Dullemond} \& {Dominik}(2005)}]{2005A&A...434..971D}
{Dullemond}, C.~P. \& {Dominik}, C. 2005, \aap, 434, 971

\bibitem[{{Dullemond} {et~al.}(2001){Dullemond}, {Dominik}, \&
  {Natta}}]{2001ApJ...560..957D}
{Dullemond}, C.~P., {Dominik}, C., \& {Natta}, A. 2001, \apj, 560, 957

\bibitem[{{Eulderink} \& {Mellema}(1995)}]{1995A&AS..110..587E}
{Eulderink}, F. \& {Mellema}, G. 1995, \aaps, 110, 587

\bibitem[{{Goldreich} \& {Sari}(2003)}]{2003ApJ...585.1024G}
{Goldreich}, P. \& {Sari}, R. 2003, \apj, 585, 1024

\bibitem[{{Goldreich} \& {Tremaine}(1980)}]{1980ApJ...241..425G}
{Goldreich}, P. \& {Tremaine}, S. 1980, \apj, 241, 425

\bibitem[{{Hayes} \& {Norman}(2003)}]{2003ApJS..147..197H}
{Hayes}, J.~C. \& {Norman}, M.~L. 2003, \apjs, 147, 197

\bibitem[{{Ida} \& {Lin}(2004{\natexlab{a}})}]{2004ApJ...604..388I}
{Ida}, S. \& {Lin}, D.~N.~C. 2004{\natexlab{a}}, \apj, 604, 388

\bibitem[{{Ida} \& {Lin}(2004{\natexlab{b}})}]{2004ApJ...616..567I}
{Ida}, S. \& {Lin}, D.~N.~C. 2004{\natexlab{b}}, \apj, 616, 567

\bibitem[{{Ida} \& {Lin}(2005)}]{2005ApJ...626.1045I}
{Ida}, S. \& {Lin}, D.~N.~C. 2005, \apj, 626, 1045

\bibitem[{{Jang-Condell} \& {Sasselov}(2003)}]{2003ApJ...593.1116J}
{Jang-Condell}, H. \& {Sasselov}, D.~D. 2003, \apj, 593, 1116

\bibitem[{{Jang-Condell} \& {Sasselov}(2004)}]{2004ApJ...608..497J}
{Jang-Condell}, H. \& {Sasselov}, D.~D. 2004, \apj, 608, 497

\bibitem[{{Jang-Condell} \& {Sasselov}(2005)}]{2005ApJ...619.1123J}
{Jang-Condell}, H. \& {Sasselov}, D.~D. 2005, \apj, 619, 1123

\bibitem[{{Klahr} \& {Kley}(2006)}]{2006A&A...445..747K}
{Klahr}, H. \& {Kley}, W. 2006, \aap, 445, 747

\bibitem[{{Klahr} \& {Bodenheimer}(2003)}]{2003ApJ...582..869K}
{Klahr}, H.~H. \& {Bodenheimer}, P. 2003, \apj, 582, 869

\bibitem[{{Kley}(1989)}]{1989A&A...208...98K}
{Kley}, W. 1989, \aap, 208, 98

\bibitem[{{Kley}(1998)}]{1998A&A...338L..37K}
{Kley}, W. 1998, \aap, 338, L37

\bibitem[{{Kley}(1999)}]{1999MNRAS.303..696K}
{Kley}, W. 1999, \mnras, 303, 696

\bibitem[{{Kley} \& {Dirksen}(2006)}]{2006A&A...447..369K}
{Kley}, W. \& {Dirksen}, G. 2006, \aap, 447, 369

\bibitem[{{Levermore} \& {Pomraning}(1981)}]{1981ApJ...248..321L}
{Levermore}, C.~D. \& {Pomraning}, G.~C. 1981, \apj, 248, 321

\bibitem[{{Lubow} {et~al.}(1999){Lubow}, {Seibert}, \&
  {Artymowicz}}]{1999ApJ...526.1001L}
{Lubow}, S.~H., {Seibert}, M., \& {Artymowicz}, P. 1999, \apj, 526, 1001

\bibitem[{{Masset} {et~al.}(2006){Masset}, {D'Angelo}, \&
  {Kley}}]{2006ApJ...652..730M}
{Masset}, F.~S., {D'Angelo}, G., \& {Kley}, W. 2006, \apj, 652, 730

\bibitem[{{Masset} \& {Papaloizou}(2003)}]{2003ApJ...588..494M}
{Masset}, F.~S. \& {Papaloizou}, J.~C.~B. 2003, \apj, 588, 494

\bibitem[{{Mellema} {et~al.}(1991){Mellema}, {Eulderink}, \&
  {Icke}}]{1991A&A...252..718M}
{Mellema}, G., {Eulderink}, F., \& {Icke}, V. 1991, \aap, 252, 718

\bibitem[{{Menou} \& {Goodman}(2004)}]{2004ApJ...606..520M}
{Menou}, K. \& {Goodman}, J. 2004, \apj, 606, 520

\bibitem[{{Morohoshi} \& {Tanaka}(2003)}]{2003MNRAS.346..915M}
{Morohoshi}, K. \& {Tanaka}, H. 2003, \mnras, 346, 915

\bibitem[{{Nelson} \& {Benz}(2003{\natexlab{a}})}]{2003ApJ...589..556N}
{Nelson}, A.~F. \& {Benz}, W. 2003{\natexlab{a}}, \apj, 589, 556

\bibitem[{{Nelson} \& {Benz}(2003{\natexlab{b}})}]{2003ApJ...589..578N}
{Nelson}, A.~F. \& {Benz}, W. 2003{\natexlab{b}}, \apj, 589, 578

\bibitem[{{Nelson} \& {Papaloizou}(2004)}]{2004MNRAS.350..849N}
{Nelson}, R.~P. \& {Papaloizou}, J.~C.~B. 2004, \mnras, 350, 849

\bibitem[{{Ogilvie} \& {Lubow}(2003)}]{2003ApJ...587..398O}
{Ogilvie}, G.~I. \& {Lubow}, S.~H. 2003, \apj, 587, 398

\bibitem[{{Paardekooper} \& {Mellema}(2006{\natexlab{a}})}]{radlett}
{Paardekooper}, S.-J. \& {Mellema}, G. 2006{\natexlab{a}}, \aap, 459, L17

\bibitem[{{Paardekooper} \&
  {Mellema}(2006{\natexlab{b}})}]{2006A&A...450.1203P}
{Paardekooper}, S.-J. \& {Mellema}, G. 2006{\natexlab{b}}, \aap, 450, 1203

\bibitem[{{Paardekooper} \& {Papaloizou}(2007)}]{subm}
{Paardekooper}, S.-J. \& {Papaloizou}, J.~C.~B. 2007, \aap, submitted

\bibitem[{{Papaloizou} \& {Nelson}(2003)}]{2003MNRAS.339..983P}
{Papaloizou}, J.~C.~B. \& {Nelson}, R.~P. 2003, \mnras, 339, 983

\bibitem[{{Papaloizou} {et~al.}(2001){Papaloizou}, {Nelson}, \&
  {Masset}}]{2001A&A...366..263P}
{Papaloizou}, J.~C.~B., {Nelson}, R.~P., \& {Masset}, F. 2001, \aap, 366, 263

\bibitem[{{Papaloizou} \& {Terquem}(1999)}]{1999ApJ...521..823P}
{Papaloizou}, J.~C.~B. \& {Terquem}, C. 1999, \apj, 521, 823

\bibitem[{{Pollack} {et~al.}(1996){Pollack}, {Hubickyj}, {Bodenheimer},
  {Lissauer}, {Podolak}, \& {Greenzweig}}]{1996Icar..124...62P}
{Pollack}, J.~B., {Hubickyj}, O., {Bodenheimer}, P., {et~al.} 1996, Icarus,
  124, 62

\bibitem[{{Roe}(1981)}]{1981...............}
{Roe}, P.~L. 1981, J.~Comp.~Phys, 43, 357

\bibitem[{{Sari} \& {Goldreich}(2004)}]{2004ApJ...606L..77S}
{Sari}, R. \& {Goldreich}, P. 2004, \apjl, 606, L77

\bibitem[{{Shakura} \& {Sunyaev}(1973)}]{1973A&A....24..337S}
{Shakura}, N.~I. \& {Sunyaev}, R.~A. 1973, \aap, 24, 337

\bibitem[{{Stone} {et~al.}(1992){Stone}, {Mihalas}, \&
  {Norman}}]{1992ApJS...80..819S}
{Stone}, J.~M., {Mihalas}, D., \& {Norman}, M.~L. 1992, \apjs, 80, 819

\bibitem[{{Tanaka} {et~al.}(2002){Tanaka}, {Takeuchi}, \&
  {Ward}}]{2002ApJ...565.1257T}
{Tanaka}, H., {Takeuchi}, T., \& {Ward}, W.~R. 2002, \apj, 565, 1257

\bibitem[{{Terquem}(2003)}]{2003MNRAS.341.1157T}
{Terquem}, C.~E.~J.~M.~L.~J. 2003, \mnras, 341, 1157

\bibitem[{{Ward}(1991)}]{1991LPI....22.1463W}
{Ward}, W.~R. 1991, in Lunar and Planetary Institute Conference Abstracts,
  Vol.~22, Lunar and Planetary Institute Conference Abstracts, 1463

\bibitem[{{Ward}(1997)}]{1997Icar..126..261W}
{Ward}, W.~R. 1997, Icarus, 126, 261

\end{thebibliography}

\end{document}